\documentclass[epsfigure,showpacs,notitlepage,onecolumn,superscriptaddress,longbibliography,floatfix]{revtex4-1}
\usepackage{dcolumn}    
\usepackage{bm} 
\usepackage{graphicx}
\usepackage{amsmath}    
\usepackage{latexsym}
\usepackage{amsfonts}   
\usepackage{amssymb}
\usepackage{array}      
\usepackage{epsfig}
\usepackage{txfonts}
\usepackage{color}
\usepackage[colorlinks=true,linkcolor=blue,urlcolor=blue,citecolor=blue]{hyperref}
\usepackage{hyperref}
\usepackage{epstopdf}
\usepackage{dsfont}

\begin{document}
 \title{Finite-time two-spin quantum Otto engines: shortcuts to adiabaticity vs. irreversibility}

\author{Bar\i\c{s} \c{C}akmak}
\affiliation{College of Engineering and Natural Sciences, Bah\c{c}e\c{s}ehir University, Be\c{s}ikta\c{s}, Istanbul 34353, Turkey}
\email{baris.cakmak@eng.bau.edu.tr}

\begin{abstract}

We propose a quantum Otto cycle in a two spin-$1/2$ anisotropic XY model in a transverse external magnetic field. We first characterize the parameter regime that the working medium operates as an engine in the adiabatic regime. Then, we consider finite-time behavior of the engine with and without utilizing a shortcut to adiabaticity (STA) technique. STA schemes guarantee that the dynamics of a system follows the adiabatic path, at the expense of introducing an external control. We compare the performance of the non-adiabatic and STA engines for a fixed adiabatic efficiency but different parameters of the working medium. We observe that, for certain parameter regimes, the irreversibility, as measured by the efficiency lags, due to finite-time driving is so low that non-adiabatic engine performs quite close to the adiabatic engine, leaving the STA engine only marginally better than the non-adiabatic one. This suggests that by designing the working medium Hamiltonian one may spare the difficulty of dealing with an external control protocol.
 
\end{abstract}

\date{\today}

\maketitle

\section{Introduction}

Recent years have witnessed an ever increasing interest and a rapid development in the quest to understand the thermodynamics of out-of-equilibrium quantum systems~\cite{DeffnerCampbellBook,JPA_Goold,TJP_Asli}. This field of research, widely known as the quantum thermodynamics, blends various branches of physics such as quantum information science, quantum optics, condensed matter physics and quantum control, to name a few. In addition to its contribution to our fundamental understanding of how the concepts of classical thermodynamics can be generalized to quantum domain, quantum thermodynamics also significantly contributes in the development of new quantum technologies that take advantage of quantum systems by actively manipulating them~\cite{QTSQR,PRXQ_Deutsch,arXiv_Mukherjee}.

One of the major problems within the realm of quantum thermodynamics stems from this active manipulation. Traditionally, thermodynamic transformations are made quasi-statically, such that they are slow enough so that the subject system is kept at equilibrium at all times. An arbitrary finite time transformation would then require some thermodynamic control~\cite{EPL_Deffner}, so that one could avoid any form of irreversibility (such as entropy production) originating from fast manipulation of the system~\cite{LandiMauro,Entropy_Dann}. Techniques of shortcuts to adiabaticity (STA) are perfectly suitable for such control purposes which make sure that the subject system ends up at the adiabatic final state of the desired transformation at a finite time~\cite{RMP_STA}. Among various STA techniques, counterdiabatic driving (CD) is probably the one that attracted the highest attention~\cite{Demirplak,Demirplak2005,Demirplak2008,JPA_Berry}. CD not only ensures that the system ends up at the adiabatic final state but also ensures that it follows the adiabatic eigenstates at all times, by introducing an external control Hamiltonian. Nevertheless, this property makes CD the STA method that is energetically most costly~\cite{NJP_Belfast}.

One particular sub-field in quantum thermodynamics that quantum control proves to be useful is quantum heat engines, which are generalizations of the corresponding classical engine cycles to cases that have quantum systems as their working mediums~\cite{PRE_Quan1,PRE_Quan2,JCP_Kosloff,PRL_Kieu,EPJD_Kieu,Entropy_Kosloff}. In addition to theoretical proposals of possible implementations~\cite{PRL_Obinna,Entropy_Deffner_transmon,PRE_MustecapliogluExp}, there are actual experiments demonstrating operational quantum heat engines~\cite{Science_Obinna,PRL_Peterson,PRL_negativetemp,PRL_Klatzow,PRA_Deng}. However, presence of adiabatic processes in the engine cycle forces them to be performed slowly, resulting in a vanishingly small power output. STA techniques have shown to be quite useful in speeding up these adiabatic processes~\cite{PRA_Deng,PRE_Gong1,SciRep_delCampo,Entropy_Beau,EPL_Obinna, PRE_Obinna, PRE_Obinna2,PRE_Baris,NJP_Steve,SciAdv_Deng,delcampo2018}, yet they come at the expense of a certain energetic cost (see also~\cite{Entropy_Kosloff,Entropy_Dann} for an alternative view). In this work, we consider a quantum Otto engine that has two coupled spins as a working medium. Following its characterization in the adiabatic limit, we investigate its finite time behavior with and without utilizing a specific STA scheme, and compare their performances in different parameter regimes. We observe that, for a fixed adiabatic efficiency, it is possible to find a set of parameters for which the engine cycle without any external control perform very close to the one with STA. We then argue that such increase in the performance is due to the reduced irreversibility due to finite time driving, as quantified by the efficiency lags (based on non-equilibrium lags)~\cite{PRL_Peterson,LandiMauro}, in these parameter regimes. 

This paper is organized as follows. In Sec.~\ref{prelim} we introduce the concepts that are central to this work such as the CD scheme, details of the quantum Otto cycle and how to characterize its performance with and without the presence of a STA scheme. We present our model for the working substance of our Otto engine in Sec.~\ref{sec:twospin}, which is two-spin anisotropic XY model in transverse magnetic field. We continue this section by identifying the parameter region in which the working substance operates as an engine and compare its performance in adiabatic, non-adiabatic and STA cases. We conclude in Sec.~\ref{sec:conclusion}.

%---------------------------------------------------------------------------------------------------------------------------------------------------------------------------------------
\section{Preliminaries}\label{prelim}
%---------------------------------------------------------------------------------------------------------------------------------------------------------------------------------------
\subsection{Counterdiabatic driving}\label{sec:ca_drive}

Assume that we have a system that is described by a time-dependent Hamiltonian $H_0(t)$. In order for this system to follow the adiabatic eigenstates of its bare Hamiltonian, any change in $H_0(t)$ must be made very slowly, in a scale set by the energy gap of the Hamiltonian. Any fast driving will induce transitions between its energy levels, resulting in the deviation from the adiabatic path. Utilizing the CD scheme~\cite{Demirplak,JPA_Berry}, it is possible to mimic the adiabatic evolution at a finite time by introducing an additional Hamiltonian, $H_{\text{CD}}(t)$, such that the system evolving with $H(t)=H_0(t)+H_{\text{CD}}(t)$ follows the adiabatic eigenstates at all times. Exact form of this Hamiltonian is given as~\cite{JPA_Berry}
\begin{equation}
H_{\text{CD}}(t)=i\hbar\sum_n\left(\partial_t|n(t)\rangle\langle n(t)|-\langle n(t)|\partial_tn(t)\rangle |n(t)\rangle\langle n(t)|\right),
\end{equation}
where $|n(t)\rangle$ is the $n^{th}$ eigenstate of the bare Hamiltonian $H_0(t)$. The necessity to diagonalize $H_0(t)$ to determine the $H_{\text{CD}}(t)$ is a demanding task for systems with large Hilbert space. Therefore, alternative approaches such as the ones that do not require the knowledge of full spectrum~\cite{PRL_delCampo,PRX_Deffner} or approximate driving schemes~\cite{PNAS_Sels} have been developed and utilized in many-body systems~\cite{PRR_Wolfgang}.

\subsection{Quantum Otto Cycle}\label{sec:cycle}

\begin{figure}[t]
\center
\includegraphics[width=0.8\columnwidth]{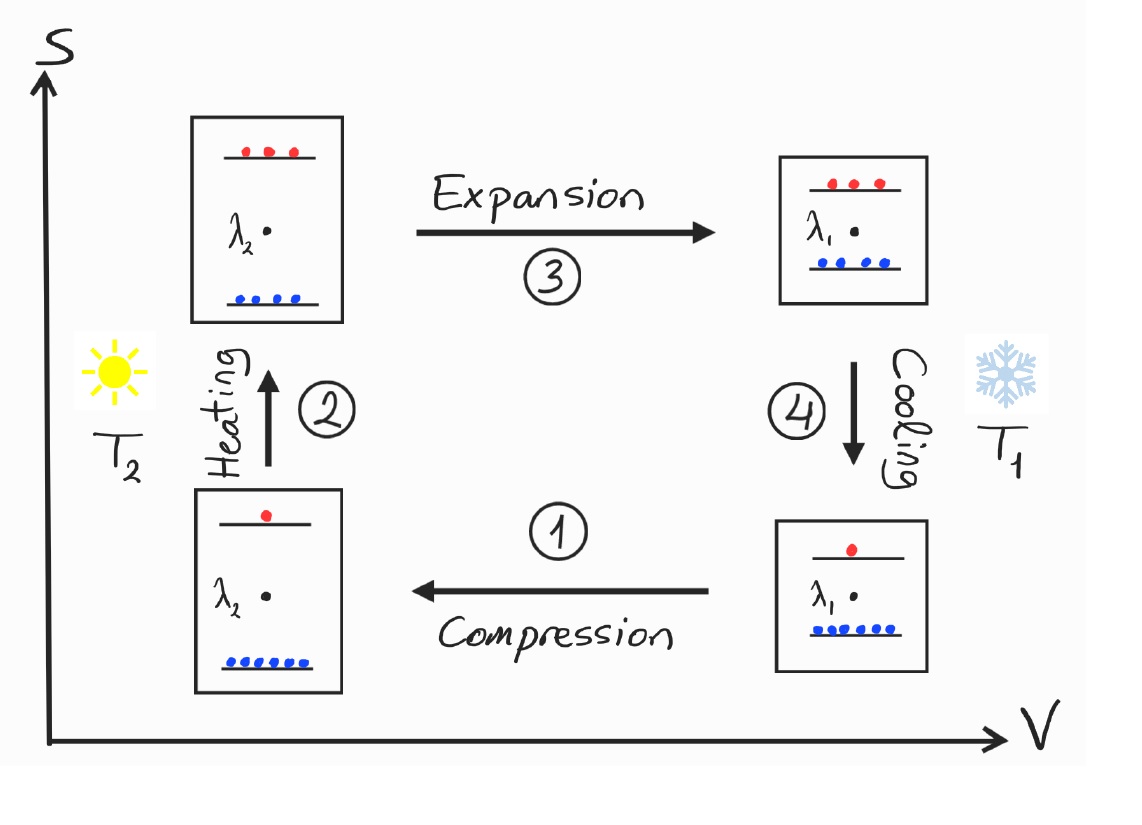}\\
\caption{Schematic representation of an ideal quantum Otto cycle in entropy (S) vs. volume (V) plane. Compression and expansion strokes are realized by changing the control parameter in the Hamiltonian of the working medium $\lambda$. When performed adiabatically, the entropy remains constant during these processes, as depicted by constant populations of the energy levels. The working medium is put in contact with a hot and cold bath during heating and cooling strokes, respectively, resulting a change in the populations, and hence in the entropy. Further details of each stroke are explained in the text.}
\label{schematics}
\end{figure}

The quantum Otto cycle is a four-stroke process, which is simply generalization of the corresponding classical cycle to quantum systems \cite{PRE_Quan1,PRE_Quan2,JCP_Kosloff,PRL_Kieu,EPJD_Kieu}. Two out of the four-strokes are adiabatic compression and expansion branches and performed by varying a control parameter, $\lambda_t$, in the Hamiltonian of the working substance, $H(\lambda_t)$. Remaining two are isochoric heating and cooling branches which involves thermalization of the working substance to the corresponding bath temperature. The cycle is schematically depicted in Fig.~\ref{schematics}, and below we briefly present the details of each stroke of the cycle:
\begin{enumerate}
\setlength\itemsep{-0.76em}
\item \textit{Compression:} The working substance is in a thermal state at the temperature of the cold bath $\rho_{\beta_1}=\exp(-\beta_1H(\lambda_1))/Z(\beta_1, \lambda_1)$, where $Z(\beta_1, \lambda_1)$ is the partition function with $\beta_1=1/k_BT_1$, and completely isolated from any heat bath. The time-dependent parameter in its Hamiltonian is increased from $\lambda_1$ to $\lambda_2$ in a time interval of $\tau_1$, resulting an increase in the internal energy of the system. During this stroke the time evolution of the working substance is unitary and ideally assumed to be slow enough so that there are no unwanted transitions between the energy levels of the system so that their populations, therefore the entropy, remain constant, in accordance with the quantum adiabatic theorem. Such an evolution guarantees that all the internal energy change of the working substance is due to the work done on it, $\langle W_1\rangle=\text{tr}[H(\lambda_2)\rho_{\text{com}}-H(\lambda_1)\rho_{\beta_1}]$, where $\rho_{\text{com}}$ denotes the state of the system at the end of the compression stroke. 
\newline
\item \textit{Heating:} The working substance is brought in contact with a hot bath at temperature $\beta_2=1/k_BT_2$ during a time interval $\tau_2$ that is sufficient for the system to thermalize to the bath temperature. No work is done in this branch and the change in the internal energy of the working substance is only due to the heat absorbed throughout the process, $\langle Q_2\rangle=\text{tr}[H(\lambda_2)\rho_{\beta_2}-H(\lambda_2)\rho_{\text{com}}]$, where $\rho_{\beta_2}=\exp(-\beta_2H(\lambda_2))/Z(\beta_2, \lambda_2)$ with $Z(\beta_2, \lambda_2)$ being the partition function. \newline
\item \textit{Expansion:} The working substance is again isolated from any bath and and the time-dependent parameter in its Hamiltonian is now decreased from $\lambda_2$ to $\lambda_1$ in a time interval of $\tau_3$, resulting an decrease in the internal energy of the system. Similar to the compression stroke, the evolution during this stroke is unitary and again should be performed slow enough to make sure the process remains adiabatic and entropy remains constant. The decrease in the working substance energy is extracted as work, 
$\langle W_3\rangle=\text{tr}[H(\lambda_1)\rho_{\text{exp}}-H(\lambda_2)\rho_{\beta_2}]$, where $\rho_{\text{exp}}$ is the state of the system at the end of the expansion stroke.  \newline
\item \textit{Cooling:} In this final branch of the cycle, the working substance is brought in contact with a cold bath at temperature $\beta_1=1/k_BT_1$ during a time interval $\tau_4$ that is sufficient for the system to thermalize to the bath temperature. No work is done in this branch and the change in the internal energy of the working substance is only due to the released heat throughout the process, $\langle Q_4\rangle=\text{tr}[H(\lambda_1)\rho_{\beta_1}-H(\lambda_1)\rho_{\text{exp}}]$.
\end{enumerate}
The unitary evolution in the compression and expansion strokes are governed by the von Neumann equation $\dot{\rho}(t)=-\frac{i}{\hbar}[H(\lambda_t), \rho(t)]$. Note that negative (positive) values of work or heat correspond to the case of these quantities being extracted from (absorbed by) the system. Therefore, in the engine cycle described above we have $\langle W_1\rangle, \langle Q_2\rangle> 0$ and $\langle W_3\rangle, \langle Q_4\rangle< 0$, such that the total work in the adiabatic case $\langle W^{A}\rangle=\langle W_1\rangle+\langle W_3\rangle<0$ implying that we have net extracted work from the working medium.

We would like to highlight that the highest performance out of the above cycle in terms of the work output and efficiency, can only be obtained if the compression and expansion strokes are performed adiabatically, as described. However, this condition can only be met if $\lambda_t$ is varied extremely slowly, which in turn results in a vanishing power output from the engine due to very long cycle times. One central aim of this manuscript is to present a way to overcome this difficulty by introducing a STA scheme, so that one can mimic the adiabatic dynamics of the working substance at a finite time, and thus yielding finite power. Furthermore, we will also consider finite-time driving without any control applied on the system, which will result in non-adiabatic excitations between energy levels, leading to a irreversible loss in the work output of the working substance. 

\subsection{Performance of the engine}\label{sec:performance}
Here, we would like to present the figures of merit of the engine and how we account for the energetic cost of applying the STA together with how we include them into these figures of merit. Considering a true adiabatic cycle, efficiency and power of an Otto engine are given by the usual expressions
\begin{align}
\eta_{\text{A}}&=-\frac{\langle W_{1}\rangle+\langle W_{3}\rangle}{\langle Q_2\rangle},  &   P_{\text{A}}&=-\frac{\langle W_{1}\rangle+\langle W_{3}\rangle}{\tau_{\text{cycle}}}.
\end{align}

On the other hand, if an STA scheme is employed to fasten the adiabatic strokes of the Otto cycle, due to the energetic cost of the external control, the efficiency and power of the engine are modified as follows~\cite{EPL_Obinna, PRE_Obinna, PRE_Obinna2, NJP_Steve}
\begin{equation}\label{sta_eff}
\eta_{\text{STA}}=-\frac{\langle W_1^{\text{STA}} \rangle+\langle W_3^{\text{STA}}\rangle}{\langle Q_2\rangle+V_1^{\text{CD}}+V_3^{\text{CD}}}
\end{equation}
and
\begin{equation}\label{sta_power}
P_{\text{STA}}=\frac{\langle W_1^{\text{STA}}\rangle+\langle W_3^{\text{STA}}\rangle-V_1^{\text{CD}}-V_3^{\text{CD}}}{\tau_{\text{cycle}}},
\end{equation}
where $\tau_{\text{cycle}}$ is the total cycle time of the engine and we characterize the cost as~\cite{PRE_Baris}
\begin{equation}\label{maincost}
V_i^{\text{CD}}\!=\!\int_0^{\tau}\langle \dot{H}_{\text{CD}}(t)\rangle_i dt
\end{equation}
with $i\!=\!1, 3$, which is the sum of the average of the time derivative of the CD Hamiltonian over the driving time and the expectation value is calculated using the state of the system driven by the bare Hamiltonian of the system. The reasoning and the details of the derivation of Eq.~\ref{maincost} can be found in~\cite{PRE_Baris} (especially Appendix A of the mentioned reference). We would like to note that the debate on quantifying the costs of STA schemes is still an ongoing one and the above definition is not unique~\cite{EPL_Obinna,PRE_Obinna,PRE_Obinna2,PRA_Zhang,PRA_Calzetta,PRA_Muga,PRL_Funo} (see~\cite{RMP_STA} for a comprehensive review).

Since the STA scheme enables us to mimic the adiabatic evolution, the total work output in these equations is equal to that of the adiabatic cycle $\langle W^{A}\rangle=\langle W_1^{\text{STA}}\rangle+\langle W_3^{\text{STA}}\rangle$. Therefore, in the absence of the CD Hamiltonian and the adiabatic evolution of the system, $\eta_{\text{STA}}$ reduces to $\eta_\text{A}$. As it is in general done in the literature~\cite{PRE_Obinna2}, we assume that the thermalization times are shorter than the times spent in the adiabatic strokes, and thus the total cycle time $\tau_{\text{cycle}}=\tau_1+\tau_3=2\tau$ for equal compression and expansion stroke times.

\section{Two-spin engine}\label{sec:twospin}

\begin{figure}[t]
{\bf (a)} \hskip0.49\columnwidth {\bf (b)}\\
\includegraphics[width=0.49\columnwidth]{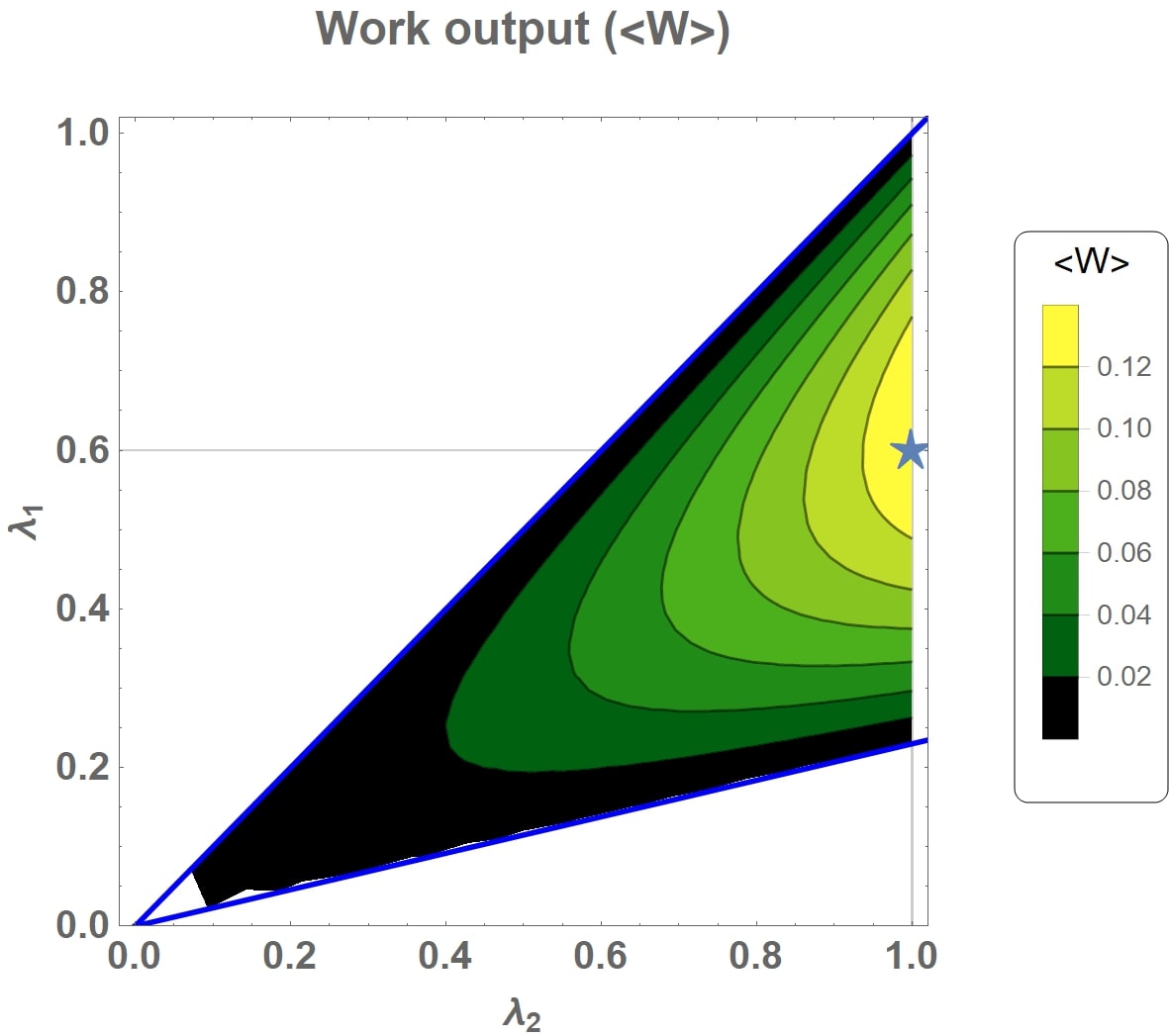}\includegraphics[width=0.49\columnwidth]{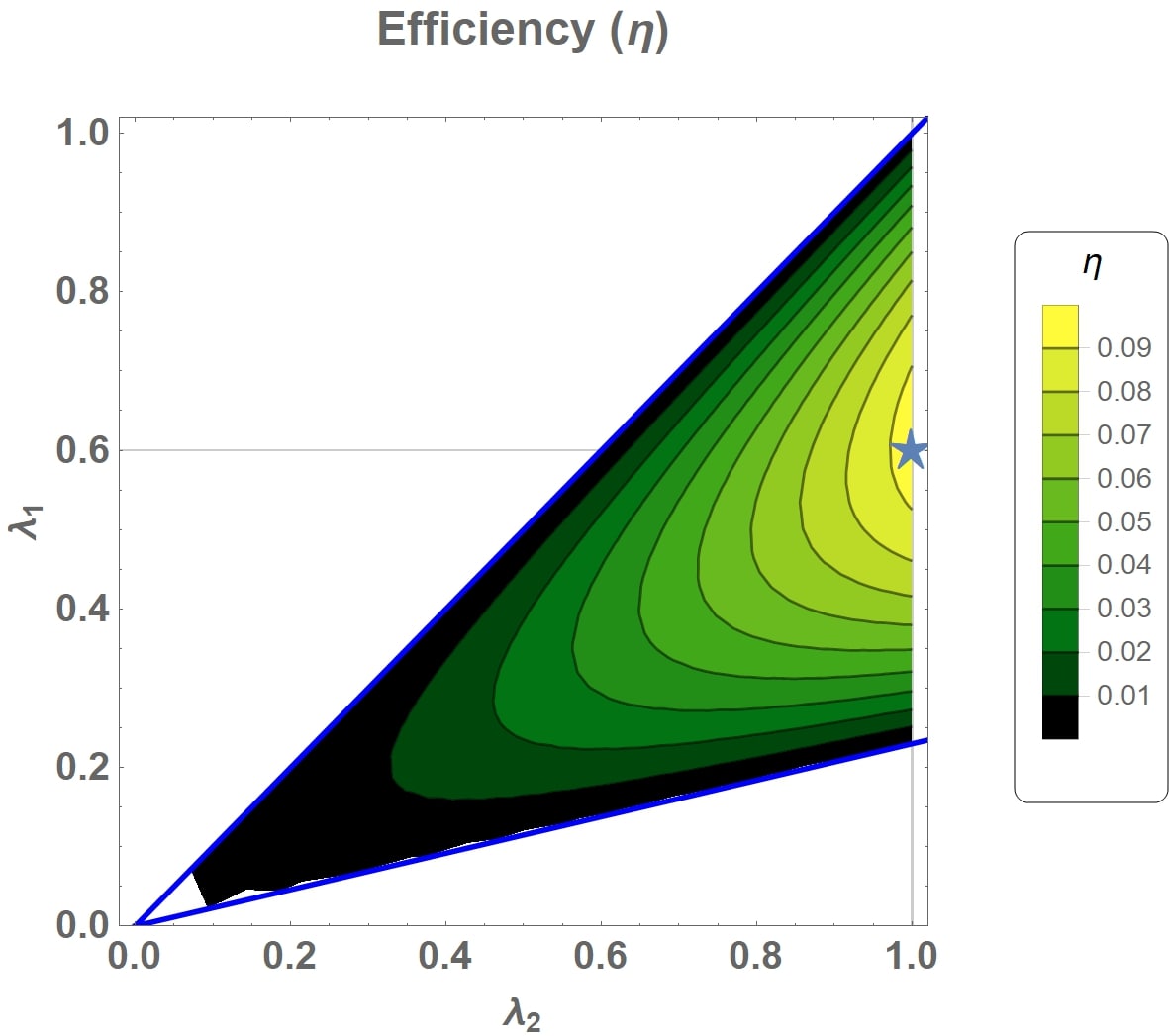}\\
\caption{The net work output {\bf (a)} and the efficiency {\bf (b)} of the two-spin engine with anisotropic XY interaction in transverse magnetic field operating between the bath temperatures $T_1\!=\!1$ and $T_2\!=\!10$. Solid blue lines denote the boundaries of the operating regimes of the engine, namely $\lambda_1=\lambda_2$ and $\lambda_1=0.23\lambda_2$. The ``$\star$" signs in both figures denote the maximum work output and the efficiency which is obtained at $\lambda_1=0.6$ and $\lambda_2=1$.}
\label{twospin}
\end{figure}

There are a number of efforts in the literature that considers coupled spin systems as working mediums of a quantum Otto engine or refrigerator~\cite{NJP_Jaramillo,PRE_Ferdi, PRE_Ferdi2, EPJP_Selcuk, PRA_Gabriele,QIP_Turkpence,PRR_Revathy}. However, there are only a few works considering a STA scheme in working substances with such composite spin systems~\cite{PRE_Baris,PRR_Wolfgang,arXiv_Wolfgang}. In what follows, we will first identify the parameter regions for which our system operates as a quantum Otto engine and then apply a STA protocol using the parameter set for which we have the maximum work output. Our working medium is composed of two-spin-$1/2$ particles and their self-Hamiltonian is characterized by an anisotropic XY model in transverse magnetic field Hamiltonian given as
\begin{equation}\label{eq:H}
H_0(t)=[1+\gamma(t)]\sigma_x^a\sigma_x^b+[1-\gamma(t)]\sigma_y^a\sigma_y^b+h(t)(\sigma_z^a+\sigma_z^b),
\end{equation}
where $\gamma(t)\in [0, 1]$ is the anisotropy parameter, $h(t)\in [0, 1]$ is the external magnetic field, $\{\sigma_x, \sigma_y, \sigma_z\}$ are the usual Pauli matrices and superscripts $a$ and $b$ are the labels for two spins. In what follows, we will suppress the explicit time-dependence of $\gamma$ and $h$ for the sake of the simplicity of the notation. The energy spectrum of Eq.~\ref{eq:H} is given as $\left\{-2, -2\sqrt{h^2+\gamma^2}, 2\sqrt{h^2+\gamma^2}, 2\right\}$. The internal energy change in the adiabatic branches of the cycle is only due to the change of the two energy levels in the middle of the spectrum.  It is possible to analytically calculate the work output and the efficiency of the two-spin engine in the adiabatic regime as follows~\cite{PRE_Baris}
\begin{equation}
\langle W^A\rangle=-\left(\langle W_1\rangle+\langle W_3\rangle\right)=\frac{2C\left(\lambda _1-\lambda _2\right)}{\left[\cosh \left(\frac{2 \lambda _1}{T_1}\right)+\cosh \left(\frac{2}{T_1}\right)\right] \left[\cosh \left(\frac{2 \lambda
   _2}{T_2}\right)+\cosh \left(\frac{2}{T_2}\right)\right]},
\end{equation}
\begin{equation}
\eta_A=\frac{C\left(\lambda _1-\lambda _2\right)}{C\lambda _2+\sinh \left(\frac{2}{T_2}\right) \cosh
   \left(\frac{2 \lambda _1}{T_1}\right)-\sinh \left(\frac{2}{T_1}\right) \cosh \left(\frac{2 \lambda _2}{T_2}\right)+\sinh \left(\frac{2}{T_2}-\frac{2}{T_1}\right)},
\end{equation}
where $C=\left\{\sinh \left(\frac{2 \lambda _2}{T_2}\right)\left[\cosh \left(\frac{2 \lambda _1}{T_1}\right)+\cosh \left(\frac{2}{T_1}\right)\right]-\sinh \left(\frac{2 \lambda _1}{T_1}\right)
   \left[\cosh \left(\frac{2 \lambda _2}{T_2}\right)+\cosh \left(\frac{2}{T_2}\right)\right]\right\}$, $\lambda_{x}=\sqrt{h_{x}+\gamma_{x}}$ with $x=1, 2$ denoting the initial and final values of the anisotropy parameter and/or the external field in steps $(i)$ and $(iii)$ of the cycle described in Sec.~\ref{sec:cycle}, $T_1$ is the temperature of the cold bath and $T_2$ is the temperature of the hot bath. It is important to note that both the extracted work and efficiency is not individually dependent on $h$ of $\gamma$, but a combination of them $\sqrt{h^2+\gamma^2}$. In Fig.~\ref{twospin}, we plot these quantities as a function of $\lambda_1$ and $\lambda_2$ and observe that it is only possible to get a working quantum Otto engine for a specific parameter regime. 
Naturally, one has to have $\lambda_1<\lambda_2$ so that we increase the internal energy in the compression stroke and, together with the energy absorbed during the heating, extract it in the expansion stroke. Interestingly however, there is also a lower bound to $\lambda_1$. Below the line $\lambda_1\approx 0.23\lambda_2$, marked by the lower thick blue line in Fig.~\ref{twospin}, the working substance again ceases to operate as an Otto engine, i.e. do not generate a net work output. Within the region that one has an operating engine, the maximum work output, and the corresponding efficiency $\eta=0.095$, is obtained for $\lambda_1=0.6$ and $\lambda_2=1$, which is marked with a ``$\star$" in Fig.s~\ref{twospin} {\bf (a)} and {\bf (b)}. 

\subsection{Finite-time operation with and without CD}

We will now attempt to accelerate adiabatic cycle with maximum work output and efficiency utilizing a STA scheme, specifically through a CD, and see if and how much we can improve the power output of the engine. For comparison, we will also analyze the performance of the engine without any control, i.e. its true finite-time, non-adiabatic behavior. Our aim in making this comparison is to see, despite the STA costs, is it meaningful, and if so, how much advantageous it is to apply a CD scheme to mimic adiabatic dynamics at finite-time considering the performance of the non-adiabatic engine.  

Our analysis will be focused around the system parameters that generate the maximum efficiency, i.e. we will fix $\lambda_1=0.6$ and $\lambda_2=1$.  We mainly have three different options in driving $\lambda$ between these values: ($i$) changing $h$ while keeping $\gamma$ constant, ($ii$) changing $\gamma$ while keeping $h$ constant, ($iii$) changing both $h$ and $\gamma$. We would like to note that the results obtained in cases ($i$) and ($ii$) have no significant qualitative difference from each other, therefore we continue with case ($i$) considering that varying the external field is simpler than controlling the interactions between the qubits. In addition, our results on case ($i$) below will prove that with the appropriate choice of system parameters it is possible operate near the highest efficiency even without any STA. As a result, we leave case ($iii$) aside since it presents an unnecessarily complicated scenario that requires simultaneous driving in both interactions and external field.

The CD term that generates the STA for the bare system Hamiltonian given in Eq.~\ref{eq:H} has the following general form~\cite{PRE_Takahashi} (see also~\cite{PRL_delCampoIsing})
\begin{equation}
H_{\text{CD}}(t)=\frac{h\dot{\gamma}-\dot{h}\gamma}{4(h^2+\gamma^2)}(\sigma_x^a\sigma_y^b+\sigma_y^a\sigma_x^b).
\label{CDterm}
\end{equation}
Natural implication of case ($i$) is $\dot{\gamma}=0$, which is important in identifying the form of Eq.~\ref{CDterm}. We also require the CD driving term to vanish at the beginning and the end of the driving, i.e. $H_{CD}(t=0, \tau)=0$ so that our system is described by its original Hamiltonian at these points. It is possible to ensure this by assuming various polynomial forms for the driven system parameter~\cite{PRL_delCampo,PRL_Ibanez,PRX_Deffner,PRE_Obinna,PRA_Stefanatos,PRE_Baris} depending on the desired smoothness of the driving. However, we have not seen any significant difference between them for our purposes in this work, and therefore continue with the simple choice below which ensures continuous first time derivatives at the boundaries
\begin{equation}
h=h_1-6(h_1-h_2)\frac{t^2}{\tau^2}\left(\frac{1}{2}-\frac{t}{3\tau}\right),
\end{equation}
where $h_1$ and $h_2$ are the initial and final external field strengths in the compression stroke.

Since $\lambda=\sqrt{h^2+\gamma^2}$ and we are operating between $\lambda_1=0.6$ and $\lambda_2=1$ to achieve highest possible efficiency, the external field is varied between the following interval $h_1=\sqrt{0.6^2-\gamma^2}$ and $h_2=\sqrt{1-\gamma^2}$. Note also that the value of the parameter $\lambda_1=0.6$ constrains the value of $\gamma\geq 0.6$ to operate within physical $h$ values, but otherwise we are free to chose $\gamma$ in the closed interval $[0, 0.6]$ while keeping the efficiency same. Such freedom allows us to investigate different working medium Hamiltonians.

Even though we have fixed the adiabatic efficiency, the short and intermediate time performance of the engine does depend on the choice of $\gamma$ and the interval within which the external field is varied through unitary performing the adiabatic branches. In addition to the change in the entropy of the working medium in the thermalization strokes, it is known that finite-time driving of a closed quantum system in the adiabatic branches results in an irreversibility in the system that can not be traced back to heat exchange and lead to the introduction of irreversible work, $W_{\text{irr}}$~\cite{PRL_Deffner,PRL_Plastina,PRE_Feldmann,NJP_Alecce,EPJD_Selcuk} (especially see~\cite{LandiMauro} for a comprehensive review). In the context of quantum heat engines any deviation from adiabaticity in the compression and expansion strokes results in the loss of useful work and thus called inner friction~\cite{PRE_Feldmann,PRE_Feldmann2,NJP_Alecce,EPJD_Selcuk,NJP_Rezek}. In order to quantify the overall effect entropy production throughout the cycle, including those due to finite-time, non-equilibrium operation, on the efficiency of our engine, we will adopt the so-called efficiency lags~\cite{PRL_Peterson}, defined through the relation $\eta=\eta_{\text{Carnot}}-\mathcal{L}$, and the explicit form of the lag is given as 
\begin{equation}
\mathcal{L}=\frac{D\left(\rho_{\text{exp}}(t)||\rho_{\beta_2}\right)+D\left(\rho_{\text{com}}(t)||\rho_{\beta_1}\right)}{\beta_2\langle Q_1\rangle},
\label{eq:lag}
\end{equation}
where $D(\rho||\sigma)=\text{tr}[\rho\ln\rho-\rho\ln\sigma]$ is the relative entropy and $\eta_{\text{Carnot}}=1-T_1/T_2$. In fact, each term appearing in the numerator of Eq.~\ref{eq:lag} are also called as non-equilibrium lags and actually equal to the $\beta W_{\text{irr}}$ in the compression and expansion processes, respectively. We would like to direct the interested reader to Ref.~\cite{LandiMauro}, especially Sec. III-F. A similar approach have been taken in quantifying the deviation from reversibility using efficiency lags in an irreversible quantum Carnot cycle~\cite{Ferdi_lags}.

\begin{figure}[t]
{\bf (a)} \hskip0.45\columnwidth {\bf (b)} \\ 
\includegraphics[width=0.45\columnwidth]{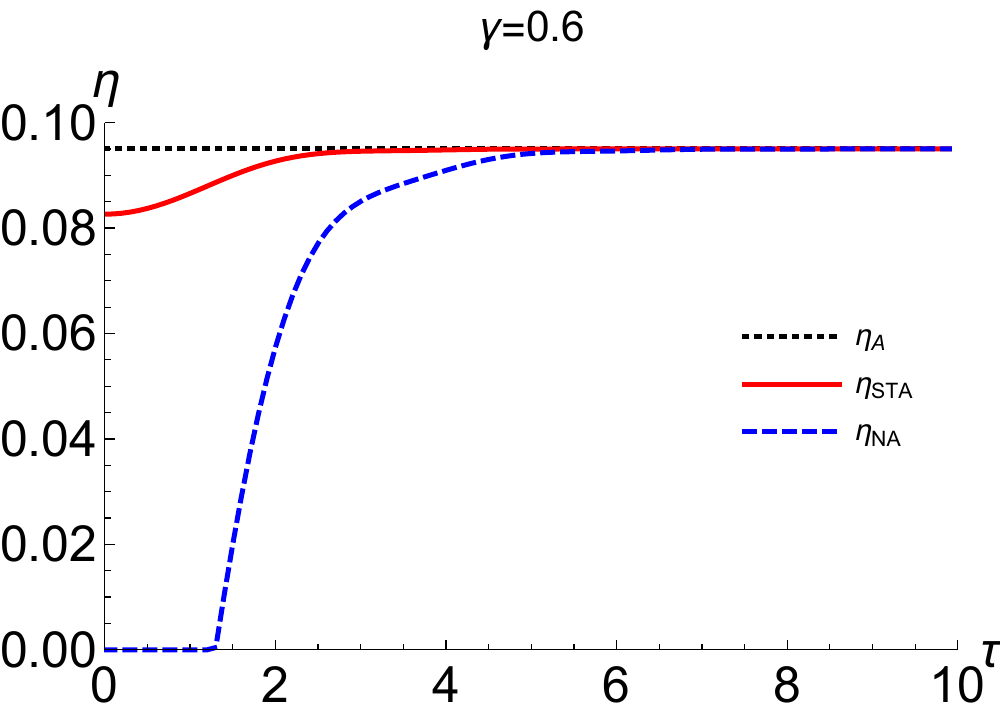}~~\includegraphics[width=0.45\columnwidth]{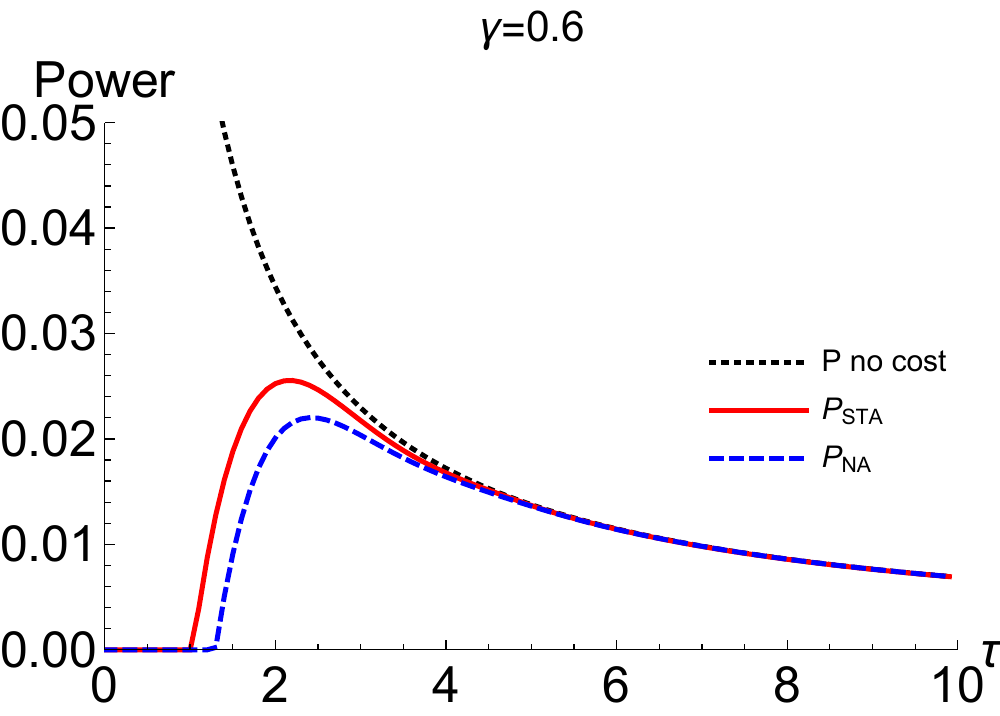} \\
{\bf (c)} \hskip0.45\columnwidth {\bf (d)}\\
\includegraphics[width=0.45\columnwidth]{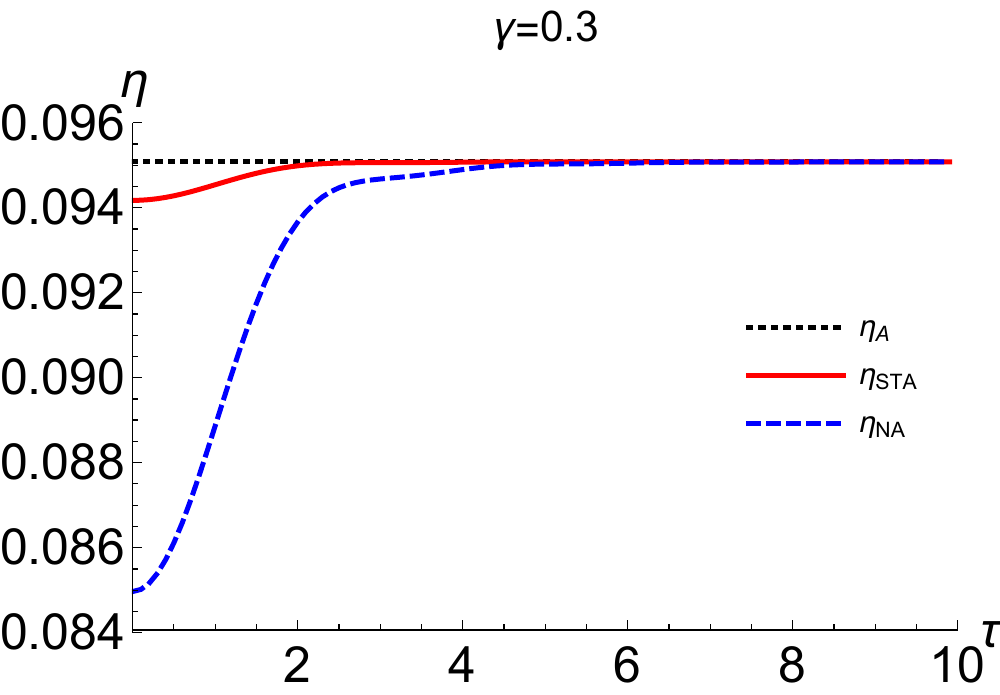}~~\includegraphics[width=0.45\columnwidth]{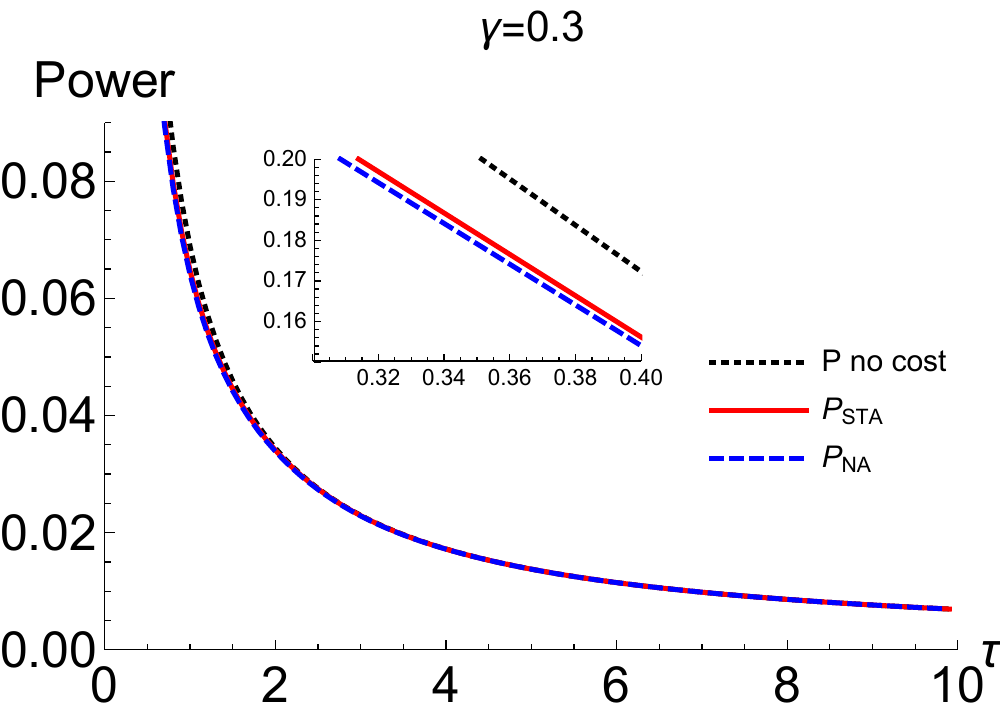}\\
\caption{Efficiency ${\bf (a)}$ and power ${\bf (b)}$ of the two-spin engine with anisotropic XY interaction in transverse magnetic field operating between the bath temperatures $T_1\!=\!1$ and $T_2\!=\!10$ with $\gamma=0.6$, $h_1=0$ and $h_2=0.8$. ${\bf (c)}$ and ${\bf (d)}$ as for the previous panels but with $\gamma=0.3$, $h_1=0.52$ to $h_2=0.95$. The inset in panel ${\bf (d)}$ displays a zoom into the curves for a better display of the differences between the presented cases. Dashed horizontal lines in ${\bf (a)}$ and ${\bf (c)}$ marks the adiabatic efficiency.}
\label{finitetime}
\end{figure}

We are now ready to compare the performances of the non-adiabatic and CD engines for two different values of $\gamma$, namely $\gamma=0.6$ and $\gamma=0.3$. While the former results in changing $h_1=0$ to $h_2=0.8$, the latter implies variation from $h_1=0.52$ to $h_2=0.95$. Note that for $\gamma=0.3$ we are required to sweep a smaller range of the external field as compared to the case of $\gamma=0.6$, which will prove to be important in the finite-time behavior of the engine. 

\begin{figure}[t]
{\bf (a)} \hskip0.48\columnwidth {\bf (b)} \\
\includegraphics[width=0.48\columnwidth]{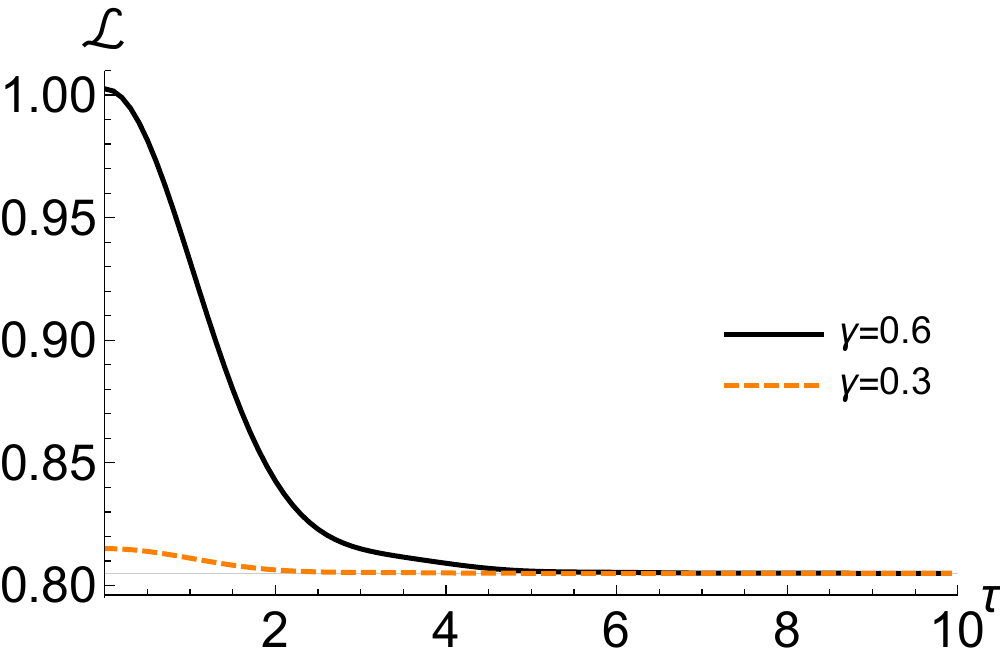}~~\includegraphics[width=0.48\columnwidth]{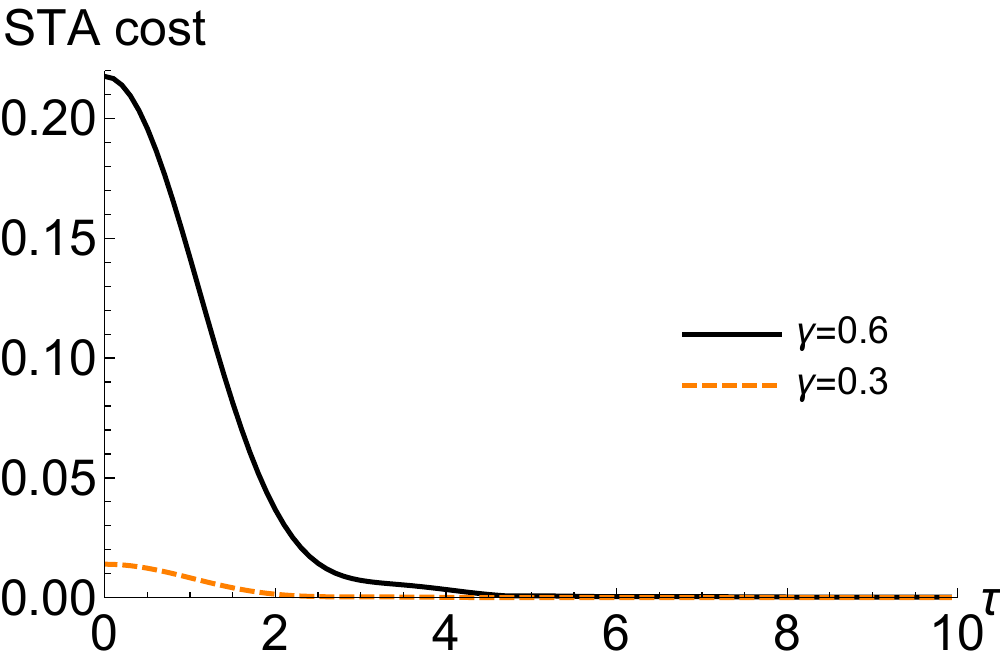}\\
\caption{Efficiency lag ${\bf (a)}$ and total STA cost ${\bf (b)}$ of the two-spin engine with anisotropic XY interaction in transverse magnetic field operating between the bath temperatures $T_1\!=\!1$ and $T_2\!=\!10$ with $\gamma=0.6$ (thick black), $\gamma=0.3$ (dashed orange).}
\label{lag_cost}
\end{figure}

In Fig.~\ref{finitetime} present our results on efficiency and power delivered by the quantum Otto engine for $\gamma=0.6$ in ${\bf (a)}-{\bf (b)}$ and for $\gamma=0.3$ in ${\bf (c)}-{\bf (d)}$, respectively, as a function of the driving time. To begin with, we observe a clear difference between the efficiencies of the engines. The non-adiabatic case for $\gamma=0.6$ fails to operate as an engine below certain driving times since it is unable to deliver work due to the irreversible excitations in the working medium induced by the fast driving. On the other hand, as displayed in Fig.~\ref{finitetime}-${\bf (b)}$, the non-adiabatic engine for $\gamma=0.3$ operates with an efficiency fairly close to that of the adiabatic engine in driving times as short as $\tau=0.001$. In both cases CD engines work with better efficiency as compared to the non-adiabatic engines since the additional term $H_{CD}(t)$ ensures that the working medium evolution is adiabatic regardless of the driving time, suppressing irreversible entropy production originating from unwanted transitions between the energy levels. The deviation of $\eta_{STA}$ from the adiabatic efficiency, $\eta_A$, stems from the energetic cost of applying the CD term that we presented in Sec.~\ref{sec:performance}. Note that the STA costs are higher in the case of $\gamma=0.6$ as compared to $\gamma=0.3$, which is related to the better performance of the latter even when there is not external control is applied. We discuss the reason behind this increased performance in detail below, in relation to our results presented in Fig.~\ref{lag_cost}. Finally, despite the considerable differences in the efficiencies, we do not observe much difference between the non-adiabatic and CD engines for both $\gamma$, but much higher power values obtained in the case of $\gamma=0.3$. Note that the inset in Fig.~\ref{finitetime}-${\bf (d)}$ displays how tiny the difference is between the CD engine and the non-adiabatic one for the mentioned case, and together they are also very close to the power output of a hypothetical engine that achieves an adiabatic work output at a finite time with no cost.

We now would like to elaborate on the superior non-adiabatic performance of the $\gamma=0.3$ case. This better performance is in fact interesting bearing in mind that the change in the energy spectrum, i.e. the change in the energy gaps of the working medium, is the same in both cases. In general, the probability of inducing unwanted excitations due to finite time driving decreases with increasing energy gap, which is clearly not the case here. However, we believe that the reason behind the better performance of the non-adiabatic engine with $\gamma=0.3$ as compared to $\gamma=0.6$ is due to the smaller variation in the external field in the former case. In Fig.~\ref{lag_cost}-${\bf (a)}$, we calculate the efficiency lag given in Eq.~\ref{eq:lag} as a function of the driving time and clearly observe that $\mathcal{L}$ is much smaller for $\gamma=0.3$, which is in accordance with the difference in the performance between the two. The minimum of $\mathcal{L}=\eta_{\text{Carnot}}-\eta_{A}$, which is denoted by the faint horizontal line in the figure, originates from the distance between the true adiabatic states at the end of expansion and compression strokes to the hot and cold baths, respectively. Therefore, any increase in $\mathcal{L}$ above this value is due to the irreversible entropy production, which results in irreversible work, caused by the fast driving of the system. Note that $\mathcal{L}$ remains very close to this minimum value for $\gamma=0.3$ suggesting that the $W_{\text{irr}}$ generated in the unitary branches due to deviations from adiabatic evolution is quite small. Similarly, the minimum driving time required for $\gamma=0.6$ non-adiabatic engine to acquire a non-zero efficiency can be understood from this plot. The Carnot efficiency for the cycle presented here is $\eta_{\text{Carnot}}=0.9$ which implies one must have $\mathcal{L}< 0.9$ that requires driving times larger than $\tau\approx 1.4$, in accordance with both Fig.~\ref{finitetime}-${\bf (a)}$ and Fig.~\ref{lag_cost}-${\bf (a)}$. In fact, any point in Fig.~\ref{finitetime}-${\bf (a)}$ and ${\bf (b)}$ for the non-adiabatic engine can be generated by using $\eta_{\text{Carnot}}$ and Fig.~\ref{lag_cost}-${\bf (a)}$ as pointed out above Eq.~\ref{eq:lag}.

As for the STA costs presented in Fig.~\ref{lag_cost}-${\bf (b)}$, we again see that it is significantly reduced for $\gamma=0.3$ from that of $\gamma=0.6$ in full agreement with our efficiency and power calculations in both cases. With the help of $\mathcal{L}$, it is possible to better understand this behavior. As mentioned many times before, the CD scheme aims to suppress any unwanted transitions as one changes $\lambda_t$ away from the adiabatic limit, i.e. suppress irreversible entropy production along the unitary strokes. The higher this irreversible entropy production the higher would the costs of applying the CD become. From Fig.~\ref{lag_cost}-${\bf (a)}$, we know that working medium is driven farther away from the adiabatic track in case of fast changes in $h$ for $\gamma=0.6$ as compared to $\gamma=0.3$, and thus, we have higher STA cost. The main difference between these two cases is the range within which we change the external field which gets smaller as $\gamma$ is reduced. Therefore, this lead us to conclude that the enhanced performance in the latter case stems from such restricted variation in $h$, which is in accordance with previous works~\cite{PRE_Baris}. Note again that we are able to achieve the same adiabatic efficiency with a smaller change in $h$ for small $\gamma$ is by exploiting the dependence of the energy spectrum of the working medium to these two parameters.

\begin{figure}[t]
\center
\includegraphics[width=0.6\columnwidth]{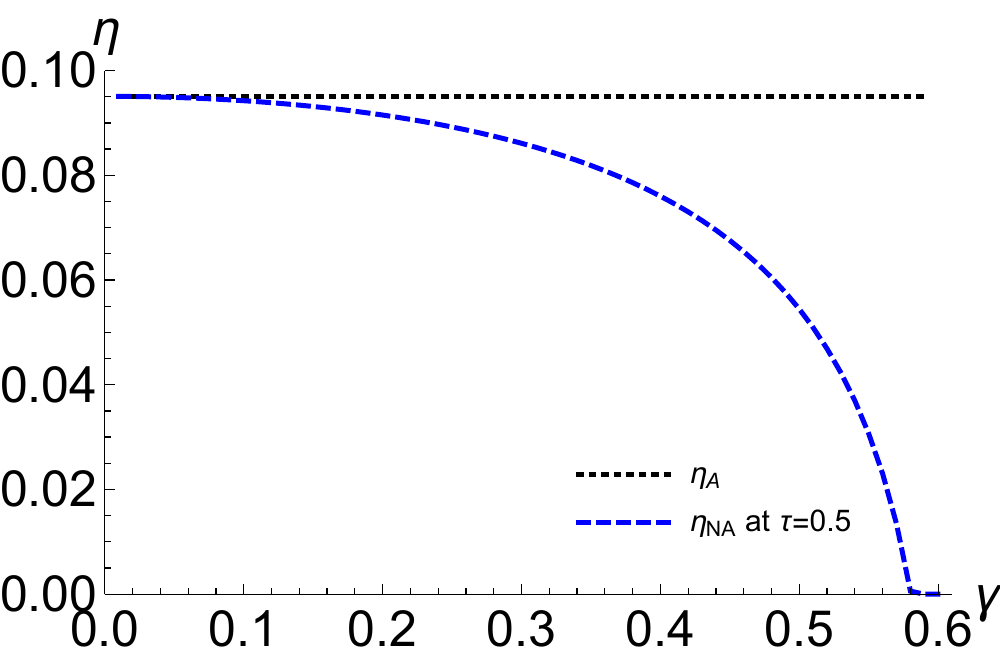}\\
\caption{Efficiency as a function of the anisotropy parameter of the two-spin engine with anisotropic XY interaction in transverse magnetic field operating between the bath temperatures $T_1\!=\!1$ and $T_2\!=\!10$ at driving time $\tau=0.5$. Dashed horizontal line marks the adiabatic efficiency.}
\label{effvsgamma}
\end{figure}

Finally, we would like to take a closer look on the dependence of the efficiency in case of the non-adiabatic engine on the anisotropy parameter $\gamma$. To that end, we fix the driving time to be $\tau=0.5$ which is clearly away from the adiabatic limit and short enough to highlight the improved performance we obtain as we lower $\gamma$. Our result is presented in Fig.~\ref{effvsgamma} and we observe that as $\gamma$ is increased, efficiency at the aforementioned driving time is quickly reduced due to irreversible entropy production caused by the fast driving. An interesting feature visible in this plot is the fact that the non-adiabatic engine efficiency converges to the adiabatic efficiency as $\gamma\rightarrow 0$, which is the isotropic limit of the model. In fact, right in this limit the CD Hamiltonian given in Eq.~\ref{CDterm} goes to zero, and the total driving Hamiltonian becomes equal to the bare Hamiltonian, $H(t)=H_0(t)$. Therefore, time evolution of the system governed by the von Neumann equation actually follows the adiabatic state and one can drive the system at an arbitrary speed without any additional control. This is called a fixed-point condition in~\cite{PRE_Takahashi} and shown to be trivial in case of a single spin but can have non-trivial consequences in more complicated systems such as the one considered in this work. In~\cite{PRE_Baris}, the authors introduced the anisotropy parameter to avoid this fixed-point, in order to make a solid analysis of the STA engine of two-spins. However, here we show that it is in fact possible to exploit this fixed-point condition in a quantum heat engine cycle to avoid irreversibility without going through the complications of the STA scheme.

\section{Conclusion and outlook}\label{sec:conclusion}

We consider a quantum Otto cycle with a working medium described by the two spin-$1/2$ anisotropic $XY$ model in a transverse magnetic field. Following the full characterization of the parameter regime for which the coupled spin system operates as an engine in the adiabatic limit and identifying the maximum efficiency, we focus on the finite-time behavior of the engine. To mimic adiabatic dynamics at a finite-time we apply a STA scheme through CD taking the energetic cost of it fully into account in the evaluation of the performance of the engine. In addition, we analyze the actual finite-time dynamics of the engine without utilizing any external control protocol and compare it with the STA performance. We observe that when we fix the efficiency of the engine to be maximal, as the anisotropy parameter is decreased the irreversibility of the non-adiabatic engine due to finite-time driving of the external field, as measured by the efficiency lags, becomes very small. This results in a significant increase in both efficiency and power of the non-adiabatic engine as compared to higher $\gamma$, which makes its performance closer to that of the STA engine. Our results suggest that for certain parameters of the Hamiltonian describing the working medium, implementing a STA scheme is not necessary and the non-adiabatic engine can operate with a similar performance due to reduced irreversibility.

Our results may have the potential to contribute to the quest of designing energy efficient quantum thermal machines. Even though STA methods are to be perfectly suitable to fasten the adiabatic strokes in a quantum heat engine cycle, they are in general resource intensive (especially CD) both on the control side and energetically~\cite{NJP_Belfast}. An alternative approach was put forward in~\cite{arXiv_Singh} in which the authors consider a two-level quantum Otto engine and refrigerator without any external control, and focus on identifying the efficiency and power of the machine by optimizing the ecological function that takes the trade-off between increased power output and entropy production into account. Building on our results that demonstrate the presence of reduced irreversibility in certain parameter regimes for two-spin quantum Otto engines, it is possible to make a more systematic analysis based on the approach of~\cite{arXiv_Singh}, which we leave as a future work. Another interesting direction could be to utilize the machine learning methods in improving the performance of quantum thermal machines. Recently in~\cite{PRL_Sgroi}, a reinforcement learning technique is introduced to reduce the entropy production in a closed quantum system due to a finite-time driving. Such an approach is perfectly suitable to be utilized in the work strokes of a quantum heat engine cycle. Specifically in the model that we have considered, one can take advantage of this method and systematically study the whole parameter landscape, learning regions of reduced entropy production.

\section*{Acknowledgment}
The author is supported by the BAGEP Award of the Science Academy and by The Research Fund of Bah\c{c}e\c{s}ehir University (BAUBAP) under project no: BAP.2019.02.03.

\bibliography{transitionless_qhe}

\begin{thebibliography}{69}%
\makeatletter
\providecommand \@ifxundefined [1]{%
 \@ifx{#1\undefined}
}%
\providecommand \@ifnum [1]{%
 \ifnum #1\expandafter \@firstoftwo
 \else \expandafter \@secondoftwo
 \fi
}%
\providecommand \@ifx [1]{%
 \ifx #1\expandafter \@firstoftwo
 \else \expandafter \@secondoftwo
 \fi
}%
\providecommand \natexlab [1]{#1}%
\providecommand \enquote  [1]{``#1''}%
\providecommand \bibnamefont  [1]{#1}%
\providecommand \bibfnamefont [1]{#1}%
\providecommand \citenamefont [1]{#1}%
\providecommand \href@noop [0]{\@secondoftwo}%
\providecommand \href [0]{\begingroup \@sanitize@url \@href}%
\providecommand \@href[1]{\@@startlink{#1}\@@href}%
\providecommand \@@href[1]{\endgroup#1\@@endlink}%
\providecommand \@sanitize@url [0]{\catcode `\\12\catcode `\$12\catcode
  `\&12\catcode `\#12\catcode `\^12\catcode `\_12\catcode `\%12\relax}%
\providecommand \@@startlink[1]{}%
\providecommand \@@endlink[0]{}%
\providecommand \url  [0]{\begingroup\@sanitize@url \@url }%
\providecommand \@url [1]{\endgroup\@href {#1}{\urlprefix }}%
\providecommand \urlprefix  [0]{URL }%
\providecommand \Eprint [0]{\href }%
\providecommand \doibase [0]{http://dx.doi.org/}%
\providecommand \selectlanguage [0]{\@gobble}%
\providecommand \bibinfo  [0]{\@secondoftwo}%
\providecommand \bibfield  [0]{\@secondoftwo}%
\providecommand \translation [1]{[#1]}%
\providecommand \BibitemOpen [0]{}%
\providecommand \bibitemStop [0]{}%
\providecommand \bibitemNoStop [0]{.\EOS\space}%
\providecommand \EOS [0]{\spacefactor3000\relax}%
\providecommand \BibitemShut  [1]{\csname bibitem#1\endcsname}%
\let\auto@bib@innerbib\@empty
%</preamble>
\bibitem [{\citenamefont {Deffner}\ and\ \citenamefont
  {Campbell}(2019)}]{DeffnerCampbellBook}%
  \BibitemOpen
  \bibfield  {author} {\bibinfo {author} {\bibfnamefont {S.}~\bibnamefont
  {Deffner}}\ and\ \bibinfo {author} {\bibfnamefont {S.}~\bibnamefont
  {Campbell}},\ }\href {\doibase 10.1088/2053-2571/ab21c6} {\emph {\bibinfo
  {title} {Quantum Thermodynamics}}}\ (\bibinfo  {publisher} {Morgan \&
  Claypool Publishers},\ \bibinfo {year} {2019})\BibitemShut {NoStop}%
\bibitem [{\citenamefont {Goold}\ \emph {et~al.}(2016)\citenamefont {Goold},
  \citenamefont {Huber}, \citenamefont {Riera}, \citenamefont {del Rio},\ and\
  \citenamefont {Skrzypczyk}}]{JPA_Goold}%
  \BibitemOpen
  \bibfield  {author} {\bibinfo {author} {\bibfnamefont {John}\ \bibnamefont
  {Goold}}, \bibinfo {author} {\bibfnamefont {Marcus}\ \bibnamefont {Huber}},
  \bibinfo {author} {\bibfnamefont {Arnau}\ \bibnamefont {Riera}}, \bibinfo
  {author} {\bibfnamefont {L{\'\i}dia}\ \bibnamefont {del Rio}}, \ and\
  \bibinfo {author} {\bibfnamefont {Paul}\ \bibnamefont {Skrzypczyk}},\
  }\bibfield  {title} {\enquote {\bibinfo {title} {The role of quantum
  information in thermodynamics---a topical review},}\ }\href
  {http://stacks.iop.org/1751-8121/49/i=14/a=143001} {\bibfield  {journal}
  {\bibinfo  {journal} {Journal of Physics A: Mathematical and Theoretical}\
  }\textbf {\bibinfo {volume} {49}},\ \bibinfo {pages} {143001} (\bibinfo
  {year} {2016})}\BibitemShut {NoStop}%
\bibitem [{\citenamefont {Tuncer}\ and\ \citenamefont
  {M\"{u}stecapl{\i}o\u{g}lu}(2020)}]{TJP_Asli}%
  \BibitemOpen
  \bibfield  {author} {\bibinfo {author} {\bibfnamefont {Asl\i}\ \bibnamefont
  {Tuncer}}\ and\ \bibinfo {author} {\bibfnamefont {\"{O}zg\"{u}r~E.}\
  \bibnamefont {M\"{u}stecapl{\i}o\u{g}lu}},\ }\bibfield  {title} {\enquote
  {\bibinfo {title} {Quantum thermodynamics and quantum coherence engines},}\
  }\href {\doibase 10.3906/fiz-2009-12} {\bibfield  {journal} {\bibinfo
  {journal} {Turk. J. Phys.}\ }\textbf {\bibinfo {volume} {44}},\ \bibinfo
  {pages} {404} (\bibinfo {year} {2020})}\BibitemShut {NoStop}%
\bibitem [{\citenamefont {MacFarlane}\ \emph {et~al.}(2003)\citenamefont
  {MacFarlane}, \citenamefont {Dowling},\ and\ \citenamefont
  {Milburn}}]{QTSQR}%
  \BibitemOpen
  \bibfield  {author} {\bibinfo {author} {\bibfnamefont {A.~G.~J.}\
  \bibnamefont {MacFarlane}}, \bibinfo {author} {\bibfnamefont {Jonathan~P.}\
  \bibnamefont {Dowling}}, \ and\ \bibinfo {author} {\bibfnamefont {Gerard~J.}\
  \bibnamefont {Milburn}},\ }\bibfield  {title} {\enquote {\bibinfo {title}
  {Quantum technology: the second quantum revolution},}\ }\href {\doibase
  10.1098/rsta.2003.1227} {\bibfield  {journal} {\bibinfo  {journal}
  {Philosophical Transactions of the Royal Society of London. Series A:
  Mathematical, Physical and Engineering Sciences}\ }\textbf {\bibinfo {volume}
  {361}},\ \bibinfo {pages} {1655--1674} (\bibinfo {year} {2003})},\ \Eprint
  {http://arxiv.org/abs/https://royalsocietypublishing.org/doi/pdf/10.1098/rsta.2003.1227}
  {https://royalsocietypublishing.org/doi/pdf/10.1098/rsta.2003.1227}
  \BibitemShut {NoStop}%
\bibitem [{\citenamefont {Deutsch}(2020)}]{PRXQ_Deutsch}%
  \BibitemOpen
  \bibfield  {author} {\bibinfo {author} {\bibfnamefont {Ivan~H.}\ \bibnamefont
  {Deutsch}},\ }\bibfield  {title} {\enquote {\bibinfo {title} {Harnessing the
  power of the second quantum revolution},}\ }\href {\doibase
  10.1103/PRXQuantum.1.020101} {\bibfield  {journal} {\bibinfo  {journal} {PRX
  Quantum}\ }\textbf {\bibinfo {volume} {1}},\ \bibinfo {pages} {020101}
  (\bibinfo {year} {2020})}\BibitemShut {NoStop}%
\bibitem [{\citenamefont {Mukherjee}\ and\ \citenamefont
  {Divakaran}(2021)}]{arXiv_Mukherjee}%
  \BibitemOpen
  \bibfield  {author} {\bibinfo {author} {\bibfnamefont {Victor}\ \bibnamefont
  {Mukherjee}}\ and\ \bibinfo {author} {\bibfnamefont {Uma}\ \bibnamefont
  {Divakaran}},\ }\bibfield  {title} {\enquote {\bibinfo {title} {Many-body
  quantum technologies},}\ }\href {https://arxiv.org/abs/2102.08301} {\bibfield
   {journal} {\bibinfo  {journal} {arXiv preprint arXiv:2102.08301}\ }
  (\bibinfo {year} {2021})}\BibitemShut {NoStop}%
\bibitem [{\citenamefont {Deffner}\ and\ \citenamefont
  {Bonan{\c{c}}a}(2020)}]{EPL_Deffner}%
  \BibitemOpen
  \bibfield  {author} {\bibinfo {author} {\bibfnamefont {Sebastian}\
  \bibnamefont {Deffner}}\ and\ \bibinfo {author} {\bibfnamefont {Marcus
  V.~S.}\ \bibnamefont {Bonan{\c{c}}a}},\ }\bibfield  {title} {\enquote
  {\bibinfo {title} {Thermodynamic control {\textemdash}an old paradigm with
  new applications},}\ }\href {\doibase 10.1209/0295-5075/131/20001} {\bibfield
   {journal} {\bibinfo  {journal} {{EPL} (Europhysics Letters)}\ }\textbf
  {\bibinfo {volume} {131}},\ \bibinfo {pages} {20001} (\bibinfo {year}
  {2020})}\BibitemShut {NoStop}%
\bibitem [{\citenamefont {Landi}\ and\ \citenamefont
  {Paternostro}(2020)}]{LandiMauro}%
  \BibitemOpen
  \bibfield  {author} {\bibinfo {author} {\bibfnamefont {Gabriel~T.}\
  \bibnamefont {Landi}}\ and\ \bibinfo {author} {\bibfnamefont {Mauro}\
  \bibnamefont {Paternostro}},\ }\bibfield  {title} {\enquote {\bibinfo {title}
  {Irreversible entropy production, from quantum to classical},}\ }\href
  {https://arxiv.org/abs/2009.07668} {\bibfield  {journal} {\bibinfo  {journal}
  {arXiv:2009.07668}\ } (\bibinfo {year} {2020})},\ \Eprint
  {http://arxiv.org/abs/2009.07668} {arXiv:2009.07668 [quant-ph]} \BibitemShut
  {NoStop}%
\bibitem [{\citenamefont {Dann}\ \emph {et~al.}(2020)\citenamefont {Dann},
  \citenamefont {Kosloff},\ and\ \citenamefont {Salamon}}]{Entropy_Dann}%
  \BibitemOpen
  \bibfield  {author} {\bibinfo {author} {\bibfnamefont {Roie}\ \bibnamefont
  {Dann}}, \bibinfo {author} {\bibfnamefont {Ronnie}\ \bibnamefont {Kosloff}},
  \ and\ \bibinfo {author} {\bibfnamefont {Peter}\ \bibnamefont {Salamon}},\
  }\bibfield  {title} {\enquote {\bibinfo {title} {Quantum finite-time
  thermodynamics: Insight from a single qubit engine},}\ }\href {\doibase
  10.3390/e22111255} {\bibfield  {journal} {\bibinfo  {journal} {Entropy}\
  }\textbf {\bibinfo {volume} {22}} (\bibinfo {year} {2020}),\
  10.3390/e22111255}\BibitemShut {NoStop}%
\bibitem [{\citenamefont {Gu\'ery-Odelin}\ \emph {et~al.}(2019)\citenamefont
  {Gu\'ery-Odelin}, \citenamefont {Ruschhaupt}, \citenamefont {Kiely},
  \citenamefont {Torrontegui}, \citenamefont {Mart\'{\i}nez-Garaot},\ and\
  \citenamefont {Muga}}]{RMP_STA}%
  \BibitemOpen
  \bibfield  {author} {\bibinfo {author} {\bibfnamefont {D.}~\bibnamefont
  {Gu\'ery-Odelin}}, \bibinfo {author} {\bibfnamefont {A.}~\bibnamefont
  {Ruschhaupt}}, \bibinfo {author} {\bibfnamefont {A.}~\bibnamefont {Kiely}},
  \bibinfo {author} {\bibfnamefont {E.}~\bibnamefont {Torrontegui}}, \bibinfo
  {author} {\bibfnamefont {S.}~\bibnamefont {Mart\'{\i}nez-Garaot}}, \ and\
  \bibinfo {author} {\bibfnamefont {J.~G.}\ \bibnamefont {Muga}},\ }\bibfield
  {title} {\enquote {\bibinfo {title} {Shortcuts to adiabaticity: Concepts,
  methods, and applications},}\ }\href {\doibase 10.1103/RevModPhys.91.045001}
  {\bibfield  {journal} {\bibinfo  {journal} {Rev. Mod. Phys.}\ }\textbf
  {\bibinfo {volume} {91}},\ \bibinfo {pages} {045001} (\bibinfo {year}
  {2019})}\BibitemShut {NoStop}%
\bibitem [{\citenamefont {Demirplak}\ and\ \citenamefont
  {Rice}(2003)}]{Demirplak}%
  \BibitemOpen
  \bibfield  {author} {\bibinfo {author} {\bibfnamefont {Mustafa}\ \bibnamefont
  {Demirplak}}\ and\ \bibinfo {author} {\bibfnamefont {Stuart~A.}\ \bibnamefont
  {Rice}},\ }\bibfield  {title} {\enquote {\bibinfo {title} {Adiabatic
  population transfer with control fields},}\ }\href {\doibase
  10.1021/jp030708a} {\bibfield  {journal} {\bibinfo  {journal} {The Journal of
  Physical Chemistry A}\ }\textbf {\bibinfo {volume} {107}},\ \bibinfo {pages}
  {9937--9945} (\bibinfo {year} {2003})},\ \Eprint
  {http://arxiv.org/abs/https://doi.org/10.1021/jp030708a}
  {https://doi.org/10.1021/jp030708a} \BibitemShut {NoStop}%
\bibitem [{\citenamefont {Demirplak}\ and\ \citenamefont
  {Rice}(2005)}]{Demirplak2005}%
  \BibitemOpen
  \bibfield  {author} {\bibinfo {author} {\bibfnamefont {Mustafa}\ \bibnamefont
  {Demirplak}}\ and\ \bibinfo {author} {\bibfnamefont {Stuart~A.}\ \bibnamefont
  {Rice}},\ }\bibfield  {title} {\enquote {\bibinfo {title} {Assisted adiabatic
  passage revisited},}\ }\href {\doibase 10.1021/jp040647w} {\bibfield
  {journal} {\bibinfo  {journal} {The Journal of Physical Chemistry B}\
  }\textbf {\bibinfo {volume} {109}},\ \bibinfo {pages} {6838--6844} (\bibinfo
  {year} {2005})},\ \bibinfo {note} {pMID: 16851769},\ \Eprint
  {http://arxiv.org/abs/https://doi.org/10.1021/jp040647w}
  {https://doi.org/10.1021/jp040647w} \BibitemShut {NoStop}%
\bibitem [{\citenamefont {Demirplak}\ and\ \citenamefont
  {Rice}(2008)}]{Demirplak2008}%
  \BibitemOpen
  \bibfield  {author} {\bibinfo {author} {\bibfnamefont {Mustafa}\ \bibnamefont
  {Demirplak}}\ and\ \bibinfo {author} {\bibfnamefont {Stuart~A.}\ \bibnamefont
  {Rice}},\ }\bibfield  {title} {\enquote {\bibinfo {title} {On the
  consistency, extremal, and global properties of counterdiabatic fields},}\
  }\href {\doibase 10.1063/1.2992152} {\bibfield  {journal} {\bibinfo
  {journal} {The Journal of Chemical Physics}\ }\textbf {\bibinfo {volume}
  {129}},\ \bibinfo {pages} {154111} (\bibinfo {year} {2008})},\ \Eprint
  {http://arxiv.org/abs/https://doi.org/10.1063/1.2992152}
  {https://doi.org/10.1063/1.2992152} \BibitemShut {NoStop}%
\bibitem [{\citenamefont {Berry}(2009)}]{JPA_Berry}%
  \BibitemOpen
  \bibfield  {author} {\bibinfo {author} {\bibfnamefont {M~V}\ \bibnamefont
  {Berry}},\ }\bibfield  {title} {\enquote {\bibinfo {title} {Transitionless
  quantum driving},}\ }\href {http://stacks.iop.org/1751-8121/42/i=36/a=365303}
  {\bibfield  {journal} {\bibinfo  {journal} {Journal of Physics A:
  Mathematical and Theoretical}\ }\textbf {\bibinfo {volume} {42}},\ \bibinfo
  {pages} {365303} (\bibinfo {year} {2009})}\BibitemShut {NoStop}%
\bibitem [{\citenamefont {Abah}\ \emph {et~al.}(2019)\citenamefont {Abah},
  \citenamefont {Puebla}, \citenamefont {Kiely}, \citenamefont {Chiara},
  \citenamefont {Paternostro},\ and\ \citenamefont {Campbell}}]{NJP_Belfast}%
  \BibitemOpen
  \bibfield  {author} {\bibinfo {author} {\bibfnamefont {Obinna}\ \bibnamefont
  {Abah}}, \bibinfo {author} {\bibfnamefont {Ricardo}\ \bibnamefont {Puebla}},
  \bibinfo {author} {\bibfnamefont {Anthony}\ \bibnamefont {Kiely}}, \bibinfo
  {author} {\bibfnamefont {Gabriele~De}\ \bibnamefont {Chiara}}, \bibinfo
  {author} {\bibfnamefont {Mauro}\ \bibnamefont {Paternostro}}, \ and\ \bibinfo
  {author} {\bibfnamefont {Steve}\ \bibnamefont {Campbell}},\ }\bibfield
  {title} {\enquote {\bibinfo {title} {Energetic cost of quantum control
  protocols},}\ }\href {\doibase 10.1088/1367-2630/ab4c8c} {\bibfield
  {journal} {\bibinfo  {journal} {New Journal of Physics}\ }\textbf {\bibinfo
  {volume} {21}},\ \bibinfo {pages} {103048} (\bibinfo {year}
  {2019})}\BibitemShut {NoStop}%
\bibitem [{\citenamefont {Quan}\ \emph {et~al.}(2007)\citenamefont {Quan},
  \citenamefont {Liu}, \citenamefont {Sun},\ and\ \citenamefont
  {Nori}}]{PRE_Quan1}%
  \BibitemOpen
  \bibfield  {author} {\bibinfo {author} {\bibfnamefont {H.~T.}\ \bibnamefont
  {Quan}}, \bibinfo {author} {\bibfnamefont {Yu-xi}\ \bibnamefont {Liu}},
  \bibinfo {author} {\bibfnamefont {C.~P.}\ \bibnamefont {Sun}}, \ and\
  \bibinfo {author} {\bibfnamefont {Franco}\ \bibnamefont {Nori}},\ }\bibfield
  {title} {\enquote {\bibinfo {title} {Quantum thermodynamic cycles and quantum
  heat engines},}\ }\href {\doibase 10.1103/PhysRevE.76.031105} {\bibfield
  {journal} {\bibinfo  {journal} {Phys. Rev. E}\ }\textbf {\bibinfo {volume}
  {76}},\ \bibinfo {pages} {031105} (\bibinfo {year} {2007})}\BibitemShut
  {NoStop}%
\bibitem [{\citenamefont {Quan}(2009)}]{PRE_Quan2}%
  \BibitemOpen
  \bibfield  {author} {\bibinfo {author} {\bibfnamefont {H.~T.}\ \bibnamefont
  {Quan}},\ }\bibfield  {title} {\enquote {\bibinfo {title} {Quantum
  thermodynamic cycles and quantum heat engines. ii.}}\ }\href {\doibase
  10.1103/PhysRevE.79.041129} {\bibfield  {journal} {\bibinfo  {journal} {Phys.
  Rev. E}\ }\textbf {\bibinfo {volume} {79}},\ \bibinfo {pages} {041129}
  (\bibinfo {year} {2009})}\BibitemShut {NoStop}%
\bibitem [{\citenamefont {Geva}\ and\ \citenamefont
  {Kosloff}(1992)}]{JCP_Kosloff}%
  \BibitemOpen
  \bibfield  {author} {\bibinfo {author} {\bibfnamefont {Eitan}\ \bibnamefont
  {Geva}}\ and\ \bibinfo {author} {\bibfnamefont {Ronnie}\ \bibnamefont
  {Kosloff}},\ }\bibfield  {title} {\enquote {\bibinfo {title} {A
  quantum‐mechanical heat engine operating in finite time. a model consisting
  of spin‐1/2 systems as the working fluid},}\ }\href {\doibase
  10.1063/1.461951} {\bibfield  {journal} {\bibinfo  {journal} {The Journal of
  Chemical Physics}\ }\textbf {\bibinfo {volume} {96}},\ \bibinfo {pages}
  {3054--3067} (\bibinfo {year} {1992})},\ \Eprint
  {http://arxiv.org/abs/https://doi.org/10.1063/1.461951}
  {https://doi.org/10.1063/1.461951} \BibitemShut {NoStop}%
\bibitem [{\citenamefont {Kieu}(2004)}]{PRL_Kieu}%
  \BibitemOpen
  \bibfield  {author} {\bibinfo {author} {\bibfnamefont {Tien~D.}\ \bibnamefont
  {Kieu}},\ }\bibfield  {title} {\enquote {\bibinfo {title} {The second law,
  maxwell's demon, and work derivable from quantum heat engines},}\ }\href
  {\doibase 10.1103/PhysRevLett.93.140403} {\bibfield  {journal} {\bibinfo
  {journal} {Phys. Rev. Lett.}\ }\textbf {\bibinfo {volume} {93}},\ \bibinfo
  {pages} {140403} (\bibinfo {year} {2004})}\BibitemShut {NoStop}%
\bibitem [{\citenamefont {Kieu}(2006)}]{EPJD_Kieu}%
  \BibitemOpen
  \bibfield  {author} {\bibinfo {author} {\bibfnamefont {T.~D.}\ \bibnamefont
  {Kieu}},\ }\bibfield  {title} {\enquote {\bibinfo {title} {Quantum heat
  engines, the second law and maxwell's daemon},}\ }\href {\doibase
  10.1140/epjd/e2006-00075-5} {\bibfield  {journal} {\bibinfo  {journal} {The
  European Physical Journal D - Atomic, Molecular, Optical and Plasma Physics}\
  }\textbf {\bibinfo {volume} {39}},\ \bibinfo {pages} {115--128} (\bibinfo
  {year} {2006})}\BibitemShut {NoStop}%
\bibitem [{\citenamefont {Kosloff}\ and\ \citenamefont
  {Rezek}(2017)}]{Entropy_Kosloff}%
  \BibitemOpen
  \bibfield  {author} {\bibinfo {author} {\bibfnamefont {Ronnie}\ \bibnamefont
  {Kosloff}}\ and\ \bibinfo {author} {\bibfnamefont {Yair}\ \bibnamefont
  {Rezek}},\ }\bibfield  {title} {\enquote {\bibinfo {title} {The quantum
  harmonic otto cycle},}\ }\href {\doibase 10.3390/e19040136} {\bibfield
  {journal} {\bibinfo  {journal} {Entropy}\ }\textbf {\bibinfo {volume} {19}}
  (\bibinfo {year} {2017}),\ 10.3390/e19040136}\BibitemShut {NoStop}%
\bibitem [{\citenamefont {Abah}\ \emph {et~al.}(2012)\citenamefont {Abah},
  \citenamefont {Ro\ss{}nagel}, \citenamefont {Jacob}, \citenamefont {Deffner},
  \citenamefont {Schmidt-Kaler}, \citenamefont {Singer},\ and\ \citenamefont
  {Lutz}}]{PRL_Obinna}%
  \BibitemOpen
  \bibfield  {author} {\bibinfo {author} {\bibfnamefont {O.}~\bibnamefont
  {Abah}}, \bibinfo {author} {\bibfnamefont {J.}~\bibnamefont {Ro\ss{}nagel}},
  \bibinfo {author} {\bibfnamefont {G.}~\bibnamefont {Jacob}}, \bibinfo
  {author} {\bibfnamefont {S.}~\bibnamefont {Deffner}}, \bibinfo {author}
  {\bibfnamefont {F.}~\bibnamefont {Schmidt-Kaler}}, \bibinfo {author}
  {\bibfnamefont {K.}~\bibnamefont {Singer}}, \ and\ \bibinfo {author}
  {\bibfnamefont {E.}~\bibnamefont {Lutz}},\ }\bibfield  {title} {\enquote
  {\bibinfo {title} {Single-ion heat engine at maximum power},}\ }\href
  {\doibase 10.1103/PhysRevLett.109.203006} {\bibfield  {journal} {\bibinfo
  {journal} {Phys. Rev. Lett.}\ }\textbf {\bibinfo {volume} {109}},\ \bibinfo
  {pages} {203006} (\bibinfo {year} {2012})}\BibitemShut {NoStop}%
\bibitem [{\citenamefont {Cherubim}\ \emph {et~al.}(2019)\citenamefont
  {Cherubim}, \citenamefont {Brito},\ and\ \citenamefont
  {Deffner}}]{Entropy_Deffner_transmon}%
  \BibitemOpen
  \bibfield  {author} {\bibinfo {author} {\bibfnamefont {Cleverson}\
  \bibnamefont {Cherubim}}, \bibinfo {author} {\bibfnamefont {Frederico}\
  \bibnamefont {Brito}}, \ and\ \bibinfo {author} {\bibfnamefont {Sebastian}\
  \bibnamefont {Deffner}},\ }\bibfield  {title} {\enquote {\bibinfo {title}
  {Non-thermal quantum engine in transmon qubits},}\ }\href {\doibase
  10.3390/e21060545} {\bibfield  {journal} {\bibinfo  {journal} {Entropy}\
  }\textbf {\bibinfo {volume} {21}} (\bibinfo {year} {2019}),\
  10.3390/e21060545}\BibitemShut {NoStop}%
\bibitem [{\citenamefont {Hardal}\ \emph {et~al.}(2017)\citenamefont {Hardal},
  \citenamefont {Aslan}, \citenamefont {Wilson},\ and\ \citenamefont
  {M\"ustecapl\ifmmode \imath \else \i \fi{}o\ifmmode~\breve{g}\else
  \u{g}\fi{}lu}}]{PRE_MustecapliogluExp}%
  \BibitemOpen
  \bibfield  {author} {\bibinfo {author} {\bibfnamefont {Ali \"U.~C.}\
  \bibnamefont {Hardal}}, \bibinfo {author} {\bibfnamefont {Nur}\ \bibnamefont
  {Aslan}}, \bibinfo {author} {\bibfnamefont {C.~M.}\ \bibnamefont {Wilson}}, \
  and\ \bibinfo {author} {\bibfnamefont {\"Ozg\"ur~E.}\ \bibnamefont
  {M\"ustecapl\ifmmode \imath \else \i \fi{}o\ifmmode~\breve{g}\else
  \u{g}\fi{}lu}},\ }\bibfield  {title} {\enquote {\bibinfo {title} {Quantum
  heat engine with coupled superconducting resonators},}\ }\href {\doibase
  10.1103/PhysRevE.96.062120} {\bibfield  {journal} {\bibinfo  {journal} {Phys.
  Rev. E}\ }\textbf {\bibinfo {volume} {96}},\ \bibinfo {pages} {062120}
  (\bibinfo {year} {2017})}\BibitemShut {NoStop}%
\bibitem [{\citenamefont {Ro{\ss}nagel}\ \emph {et~al.}(2016)\citenamefont
  {Ro{\ss}nagel}, \citenamefont {Dawkins}, \citenamefont {Tolazzi},
  \citenamefont {Abah}, \citenamefont {Lutz}, \citenamefont {Schmidt-Kaler},\
  and\ \citenamefont {Singer}}]{Science_Obinna}%
  \BibitemOpen
  \bibfield  {author} {\bibinfo {author} {\bibfnamefont {Johannes}\
  \bibnamefont {Ro{\ss}nagel}}, \bibinfo {author} {\bibfnamefont {Samuel~T.}\
  \bibnamefont {Dawkins}}, \bibinfo {author} {\bibfnamefont {Karl~N.}\
  \bibnamefont {Tolazzi}}, \bibinfo {author} {\bibfnamefont {Obinna}\
  \bibnamefont {Abah}}, \bibinfo {author} {\bibfnamefont {Eric}\ \bibnamefont
  {Lutz}}, \bibinfo {author} {\bibfnamefont {Ferdinand}\ \bibnamefont
  {Schmidt-Kaler}}, \ and\ \bibinfo {author} {\bibfnamefont {Kilian}\
  \bibnamefont {Singer}},\ }\bibfield  {title} {\enquote {\bibinfo {title} {A
  single-atom heat engine},}\ }\href {\doibase 10.1126/science.aad6320}
  {\bibfield  {journal} {\bibinfo  {journal} {Science}\ }\textbf {\bibinfo
  {volume} {352}},\ \bibinfo {pages} {325--329} (\bibinfo {year} {2016})},\
  \Eprint
  {http://arxiv.org/abs/http://science.sciencemag.org/content/352/6283/325.full.pdf}
  {http://science.sciencemag.org/content/352/6283/325.full.pdf} \BibitemShut
  {NoStop}%
\bibitem [{\citenamefont {Peterson}\ \emph {et~al.}(2019)\citenamefont
  {Peterson}, \citenamefont {Batalh\~ao}, \citenamefont {Herrera},
  \citenamefont {Souza}, \citenamefont {Sarthour}, \citenamefont {Oliveira},\
  and\ \citenamefont {Serra}}]{PRL_Peterson}%
  \BibitemOpen
  \bibfield  {author} {\bibinfo {author} {\bibfnamefont {John P.~S.}\
  \bibnamefont {Peterson}}, \bibinfo {author} {\bibfnamefont {Tiago~B.}\
  \bibnamefont {Batalh\~ao}}, \bibinfo {author} {\bibfnamefont {Marcela}\
  \bibnamefont {Herrera}}, \bibinfo {author} {\bibfnamefont {Alexandre~M.}\
  \bibnamefont {Souza}}, \bibinfo {author} {\bibfnamefont {Roberto~S.}\
  \bibnamefont {Sarthour}}, \bibinfo {author} {\bibfnamefont {Ivan~S.}\
  \bibnamefont {Oliveira}}, \ and\ \bibinfo {author} {\bibfnamefont
  {Roberto~M.}\ \bibnamefont {Serra}},\ }\bibfield  {title} {\enquote {\bibinfo
  {title} {Experimental characterization of a spin quantum heat engine},}\
  }\href {\doibase 10.1103/PhysRevLett.123.240601} {\bibfield  {journal}
  {\bibinfo  {journal} {Phys. Rev. Lett.}\ }\textbf {\bibinfo {volume} {123}},\
  \bibinfo {pages} {240601} (\bibinfo {year} {2019})}\BibitemShut {NoStop}%
\bibitem [{\citenamefont {de~Assis}\ \emph {et~al.}(2019)\citenamefont
  {de~Assis}, \citenamefont {de~Mendon\ifmmode~\mbox{\c{c}}\else \c{c}\fi{}a},
  \citenamefont {Villas-Boas}, \citenamefont {de~Souza}, \citenamefont
  {Sarthour}, \citenamefont {Oliveira},\ and\ \citenamefont
  {de~Almeida}}]{PRL_negativetemp}%
  \BibitemOpen
  \bibfield  {author} {\bibinfo {author} {\bibfnamefont {Rog\'erio~J.}\
  \bibnamefont {de~Assis}}, \bibinfo {author} {\bibfnamefont {Taysa~M.}\
  \bibnamefont {de~Mendon\ifmmode~\mbox{\c{c}}\else \c{c}\fi{}a}}, \bibinfo
  {author} {\bibfnamefont {Celso~J.}\ \bibnamefont {Villas-Boas}}, \bibinfo
  {author} {\bibfnamefont {Alexandre~M.}\ \bibnamefont {de~Souza}}, \bibinfo
  {author} {\bibfnamefont {Roberto~S.}\ \bibnamefont {Sarthour}}, \bibinfo
  {author} {\bibfnamefont {Ivan~S.}\ \bibnamefont {Oliveira}}, \ and\ \bibinfo
  {author} {\bibfnamefont {Norton~G.}\ \bibnamefont {de~Almeida}},\ }\bibfield
  {title} {\enquote {\bibinfo {title} {Efficiency of a quantum otto heat engine
  operating under a reservoir at effective negative temperatures},}\ }\href
  {\doibase 10.1103/PhysRevLett.122.240602} {\bibfield  {journal} {\bibinfo
  {journal} {Phys. Rev. Lett.}\ }\textbf {\bibinfo {volume} {122}},\ \bibinfo
  {pages} {240602} (\bibinfo {year} {2019})}\BibitemShut {NoStop}%
\bibitem [{\citenamefont {Klatzow}\ \emph {et~al.}(2019)\citenamefont
  {Klatzow}, \citenamefont {Becker}, \citenamefont {Ledingham}, \citenamefont
  {Weinzetl}, \citenamefont {Kaczmarek}, \citenamefont {Saunders},
  \citenamefont {Nunn}, \citenamefont {Walmsley}, \citenamefont {Uzdin},\ and\
  \citenamefont {Poem}}]{PRL_Klatzow}%
  \BibitemOpen
  \bibfield  {author} {\bibinfo {author} {\bibfnamefont {James}\ \bibnamefont
  {Klatzow}}, \bibinfo {author} {\bibfnamefont {Jonas~N.}\ \bibnamefont
  {Becker}}, \bibinfo {author} {\bibfnamefont {Patrick~M.}\ \bibnamefont
  {Ledingham}}, \bibinfo {author} {\bibfnamefont {Christian}\ \bibnamefont
  {Weinzetl}}, \bibinfo {author} {\bibfnamefont {Krzysztof~T.}\ \bibnamefont
  {Kaczmarek}}, \bibinfo {author} {\bibfnamefont {Dylan~J.}\ \bibnamefont
  {Saunders}}, \bibinfo {author} {\bibfnamefont {Joshua}\ \bibnamefont {Nunn}},
  \bibinfo {author} {\bibfnamefont {Ian~A.}\ \bibnamefont {Walmsley}}, \bibinfo
  {author} {\bibfnamefont {Raam}\ \bibnamefont {Uzdin}}, \ and\ \bibinfo
  {author} {\bibfnamefont {Eilon}\ \bibnamefont {Poem}},\ }\bibfield  {title}
  {\enquote {\bibinfo {title} {Experimental demonstration of quantum effects in
  the operation of microscopic heat engines},}\ }\href {\doibase
  10.1103/PhysRevLett.122.110601} {\bibfield  {journal} {\bibinfo  {journal}
  {Phys. Rev. Lett.}\ }\textbf {\bibinfo {volume} {122}},\ \bibinfo {pages}
  {110601} (\bibinfo {year} {2019})}\BibitemShut {NoStop}%
\bibitem [{\citenamefont {Deng}\ \emph
  {et~al.}(2018{\natexlab{a}})\citenamefont {Deng}, \citenamefont {Diao},
  \citenamefont {Yu}, \citenamefont {del Campo},\ and\ \citenamefont
  {Wu}}]{PRA_Deng}%
  \BibitemOpen
  \bibfield  {author} {\bibinfo {author} {\bibfnamefont {Shujin}\ \bibnamefont
  {Deng}}, \bibinfo {author} {\bibfnamefont {Pengpeng}\ \bibnamefont {Diao}},
  \bibinfo {author} {\bibfnamefont {Qianli}\ \bibnamefont {Yu}}, \bibinfo
  {author} {\bibfnamefont {Adolfo}\ \bibnamefont {del Campo}}, \ and\ \bibinfo
  {author} {\bibfnamefont {Haibin}\ \bibnamefont {Wu}},\ }\bibfield  {title}
  {\enquote {\bibinfo {title} {Shortcuts to adiabaticity in the strongly
  coupled regime: Nonadiabatic control of a unitary fermi gas},}\ }\href
  {\doibase 10.1103/PhysRevA.97.013628} {\bibfield  {journal} {\bibinfo
  {journal} {Phys. Rev. A}\ }\textbf {\bibinfo {volume} {97}},\ \bibinfo
  {pages} {013628} (\bibinfo {year} {2018}{\natexlab{a}})}\BibitemShut
  {NoStop}%
\bibitem [{\citenamefont {Deng}\ \emph {et~al.}(2013)\citenamefont {Deng},
  \citenamefont {Wang}, \citenamefont {Liu}, \citenamefont {H\"anggi},\ and\
  \citenamefont {Gong}}]{PRE_Gong1}%
  \BibitemOpen
  \bibfield  {author} {\bibinfo {author} {\bibfnamefont {Jiawen}\ \bibnamefont
  {Deng}}, \bibinfo {author} {\bibfnamefont {Qing-hai}\ \bibnamefont {Wang}},
  \bibinfo {author} {\bibfnamefont {Zhihao}\ \bibnamefont {Liu}}, \bibinfo
  {author} {\bibfnamefont {Peter}\ \bibnamefont {H\"anggi}}, \ and\ \bibinfo
  {author} {\bibfnamefont {Jiangbin}\ \bibnamefont {Gong}},\ }\bibfield
  {title} {\enquote {\bibinfo {title} {Boosting work characteristics and
  overall heat-engine performance via shortcuts to adiabaticity: Quantum and
  classical systems},}\ }\href {\doibase 10.1103/PhysRevE.88.062122} {\bibfield
   {journal} {\bibinfo  {journal} {Phys. Rev. E}\ }\textbf {\bibinfo {volume}
  {88}},\ \bibinfo {pages} {062122} (\bibinfo {year} {2013})}\BibitemShut
  {NoStop}%
\bibitem [{\citenamefont {del Campo}\ \emph {et~al.}(2014)\citenamefont {del
  Campo}, \citenamefont {Goold},\ and\ \citenamefont
  {Paternostro}}]{SciRep_delCampo}%
  \BibitemOpen
  \bibfield  {author} {\bibinfo {author} {\bibfnamefont {Adolfo}\ \bibnamefont
  {del Campo}}, \bibinfo {author} {\bibfnamefont {John}\ \bibnamefont {Goold}},
  \ and\ \bibinfo {author} {\bibfnamefont {Mauro}\ \bibnamefont
  {Paternostro}},\ }\bibfield  {title} {\enquote {\bibinfo {title} {More bang
  for your buck: Super-adiabatic quantum engines},}\ }\href {\doibase
  10.1038/srep06208} {\bibfield  {journal} {\bibinfo  {journal} {Sci. Rep.}\
  }\textbf {\bibinfo {volume} {4}},\ \bibinfo {pages} {6208} (\bibinfo {year}
  {2014})}\BibitemShut {NoStop}%
\bibitem [{\citenamefont {Beau}\ \emph {et~al.}(2016)\citenamefont {Beau},
  \citenamefont {Jaramillo},\ and\ \citenamefont {del Campo}}]{Entropy_Beau}%
  \BibitemOpen
  \bibfield  {author} {\bibinfo {author} {\bibfnamefont {Mathieu}\ \bibnamefont
  {Beau}}, \bibinfo {author} {\bibfnamefont {Juan}\ \bibnamefont {Jaramillo}},
  \ and\ \bibinfo {author} {\bibfnamefont {Adolfo}\ \bibnamefont {del Campo}},\
  }\bibfield  {title} {\enquote {\bibinfo {title} {Scaling-up quantum heat
  engines efficiently via shortcuts to adiabaticity},}\ }\href {\doibase
  10.3390/e18050168} {\bibfield  {journal} {\bibinfo  {journal} {Entropy}\
  }\textbf {\bibinfo {volume} {18}} (\bibinfo {year} {2016}),\
  10.3390/e18050168}\BibitemShut {NoStop}%
\bibitem [{\citenamefont {Abah}\ and\ \citenamefont {Lutz}(2017)}]{EPL_Obinna}%
  \BibitemOpen
  \bibfield  {author} {\bibinfo {author} {\bibfnamefont {Obinna}\ \bibnamefont
  {Abah}}\ and\ \bibinfo {author} {\bibfnamefont {Eric}\ \bibnamefont {Lutz}},\
  }\bibfield  {title} {\enquote {\bibinfo {title} {Energy efficient quantum
  machines},}\ }\href {http://stacks.iop.org/0295-5075/118/i=4/a=40005}
  {\bibfield  {journal} {\bibinfo  {journal} {EPL (Europhysics Letters)}\
  }\textbf {\bibinfo {volume} {118}},\ \bibinfo {pages} {40005} (\bibinfo
  {year} {2017})}\BibitemShut {NoStop}%
\bibitem [{\citenamefont {Abah}\ and\ \citenamefont {Lutz}(2018)}]{PRE_Obinna}%
  \BibitemOpen
  \bibfield  {author} {\bibinfo {author} {\bibfnamefont {Obinna}\ \bibnamefont
  {Abah}}\ and\ \bibinfo {author} {\bibfnamefont {Eric}\ \bibnamefont {Lutz}},\
  }\bibfield  {title} {\enquote {\bibinfo {title} {Performance of
  shortcut-to-adiabaticity quantum engines},}\ }\href {\doibase
  10.1103/PhysRevE.98.032121} {\bibfield  {journal} {\bibinfo  {journal} {Phys.
  Rev. E}\ }\textbf {\bibinfo {volume} {98}},\ \bibinfo {pages} {032121}
  (\bibinfo {year} {2018})}\BibitemShut {NoStop}%
\bibitem [{\citenamefont {Abah}\ and\ \citenamefont
  {Paternostro}(2019)}]{PRE_Obinna2}%
  \BibitemOpen
  \bibfield  {author} {\bibinfo {author} {\bibfnamefont {Obinna}\ \bibnamefont
  {Abah}}\ and\ \bibinfo {author} {\bibfnamefont {Mauro}\ \bibnamefont
  {Paternostro}},\ }\bibfield  {title} {\enquote {\bibinfo {title}
  {Shortcut-to-adiabaticity otto engine: A twist to finite-time
  thermodynamics},}\ }\href {\doibase 10.1103/PhysRevE.99.022110} {\bibfield
  {journal} {\bibinfo  {journal} {Phys. Rev. E}\ }\textbf {\bibinfo {volume}
  {99}},\ \bibinfo {pages} {022110} (\bibinfo {year} {2019})}\BibitemShut
  {NoStop}%
\bibitem [{\citenamefont {\c{C}akmak}\ and\ \citenamefont {M\"ustecapl\ifmmode
  \imath \else \i \fi{}o\ifmmode~\breve{g}\else
  \u{g}\fi{}lu}(2019)}]{PRE_Baris}%
  \BibitemOpen
  \bibfield  {author} {\bibinfo {author} {\bibfnamefont {Bar\i\c{s}}\
  \bibnamefont {\c{C}akmak}}\ and\ \bibinfo {author} {\bibfnamefont
  {\"Ozg\"ur~E.}\ \bibnamefont {M\"ustecapl\ifmmode \imath \else \i
  \fi{}o\ifmmode~\breve{g}\else \u{g}\fi{}lu}},\ }\bibfield  {title} {\enquote
  {\bibinfo {title} {Spin quantum heat engines with shortcuts to
  adiabaticity},}\ }\href {\doibase 10.1103/PhysRevE.99.032108} {\bibfield
  {journal} {\bibinfo  {journal} {Phys. Rev. E}\ }\textbf {\bibinfo {volume}
  {99}},\ \bibinfo {pages} {032108} (\bibinfo {year} {2019})}\BibitemShut
  {NoStop}%
\bibitem [{\citenamefont {Li}\ \emph {et~al.}(2018)\citenamefont {Li},
  \citenamefont {Fogarty}, \citenamefont {Campbell}, \citenamefont {Chen},\
  and\ \citenamefont {Busch}}]{NJP_Steve}%
  \BibitemOpen
  \bibfield  {author} {\bibinfo {author} {\bibfnamefont {Jing}\ \bibnamefont
  {Li}}, \bibinfo {author} {\bibfnamefont {Thom{\'a}s}\ \bibnamefont
  {Fogarty}}, \bibinfo {author} {\bibfnamefont {Steve}\ \bibnamefont
  {Campbell}}, \bibinfo {author} {\bibfnamefont {Xi}~\bibnamefont {Chen}}, \
  and\ \bibinfo {author} {\bibfnamefont {Thomas}\ \bibnamefont {Busch}},\
  }\bibfield  {title} {\enquote {\bibinfo {title} {An efficient nonlinear
  feshbach engine},}\ }\href {http://stacks.iop.org/1367-2630/20/i=1/a=015005}
  {\bibfield  {journal} {\bibinfo  {journal} {New Journal of Physics}\ }\textbf
  {\bibinfo {volume} {20}},\ \bibinfo {pages} {015005} (\bibinfo {year}
  {2018})}\BibitemShut {NoStop}%
\bibitem [{\citenamefont {Deng}\ \emph
  {et~al.}(2018{\natexlab{b}})\citenamefont {Deng}, \citenamefont {Chenu},
  \citenamefont {Diao}, \citenamefont {Li}, \citenamefont {Yu}, \citenamefont
  {Coulamy}, \citenamefont {del Campo},\ and\ \citenamefont
  {Wu}}]{SciAdv_Deng}%
  \BibitemOpen
  \bibfield  {author} {\bibinfo {author} {\bibfnamefont {Shujin}\ \bibnamefont
  {Deng}}, \bibinfo {author} {\bibfnamefont {Aur{\'e}lia}\ \bibnamefont
  {Chenu}}, \bibinfo {author} {\bibfnamefont {Pengpeng}\ \bibnamefont {Diao}},
  \bibinfo {author} {\bibfnamefont {Fang}\ \bibnamefont {Li}}, \bibinfo
  {author} {\bibfnamefont {Shi}\ \bibnamefont {Yu}}, \bibinfo {author}
  {\bibfnamefont {Ivan}\ \bibnamefont {Coulamy}}, \bibinfo {author}
  {\bibfnamefont {Adolfo}\ \bibnamefont {del Campo}}, \ and\ \bibinfo {author}
  {\bibfnamefont {Haibin}\ \bibnamefont {Wu}},\ }\bibfield  {title} {\enquote
  {\bibinfo {title} {Superadiabatic quantum friction suppression in finite-time
  thermodynamics},}\ }\href
  {http://advances.sciencemag.org/content/4/4/eaar5909} {\bibfield  {journal}
  {\bibinfo  {journal} {Science Advances}\ }\textbf {\bibinfo {volume} {4}}
  (\bibinfo {year} {2018}{\natexlab{b}})}\BibitemShut {NoStop}%
\bibitem [{\citenamefont {del Campo}\ \emph {et~al.}(2018)\citenamefont {del
  Campo}, \citenamefont {Chenu}, \citenamefont {Deng},\ and\ \citenamefont
  {Wu}}]{delcampo2018}%
  \BibitemOpen
  \bibfield  {author} {\bibinfo {author} {\bibfnamefont {Adolfo}\ \bibnamefont
  {del Campo}}, \bibinfo {author} {\bibfnamefont {Aur{\'e}lia}\ \bibnamefont
  {Chenu}}, \bibinfo {author} {\bibfnamefont {Shujin}\ \bibnamefont {Deng}}, \
  and\ \bibinfo {author} {\bibfnamefont {Haibin}\ \bibnamefont {Wu}},\
  }\enquote {\bibinfo {title} {Friction-free quantum machines},}\ in\ \href
  {\doibase 10.1007/978-3-319-99046-0_5} {\emph {\bibinfo {booktitle}
  {Thermodynamics in the Quantum Regime: Fundamental Aspects and New
  Directions}}},\ \bibinfo {editor} {edited by\ \bibinfo {editor}
  {\bibfnamefont {Felix}\ \bibnamefont {Binder}}, \bibinfo {editor}
  {\bibfnamefont {Luis~A.}\ \bibnamefont {Correa}}, \bibinfo {editor}
  {\bibfnamefont {Christian}\ \bibnamefont {Gogolin}}, \bibinfo {editor}
  {\bibfnamefont {Janet}\ \bibnamefont {Anders}}, \ and\ \bibinfo {editor}
  {\bibfnamefont {Gerardo}\ \bibnamefont {Adesso}}}\ (\bibinfo  {publisher}
  {Springer International Publishing},\ \bibinfo {address} {Cham},\ \bibinfo
  {year} {2018})\ pp.\ \bibinfo {pages} {127--148}\BibitemShut {NoStop}%
\bibitem [{\citenamefont {del Campo}(2013)}]{PRL_delCampo}%
  \BibitemOpen
  \bibfield  {author} {\bibinfo {author} {\bibfnamefont {Adolfo}\ \bibnamefont
  {del Campo}},\ }\bibfield  {title} {\enquote {\bibinfo {title} {Shortcuts to
  adiabaticity by counterdiabatic driving},}\ }\href {\doibase
  10.1103/PhysRevLett.111.100502} {\bibfield  {journal} {\bibinfo  {journal}
  {Phys. Rev. Lett.}\ }\textbf {\bibinfo {volume} {111}},\ \bibinfo {pages}
  {100502} (\bibinfo {year} {2013})}\BibitemShut {NoStop}%
\bibitem [{\citenamefont {Deffner}\ \emph {et~al.}(2014)\citenamefont
  {Deffner}, \citenamefont {Jarzynski},\ and\ \citenamefont {del
  Campo}}]{PRX_Deffner}%
  \BibitemOpen
  \bibfield  {author} {\bibinfo {author} {\bibfnamefont {Sebastian}\
  \bibnamefont {Deffner}}, \bibinfo {author} {\bibfnamefont {Christopher}\
  \bibnamefont {Jarzynski}}, \ and\ \bibinfo {author} {\bibfnamefont {Adolfo}\
  \bibnamefont {del Campo}},\ }\bibfield  {title} {\enquote {\bibinfo {title}
  {Classical and quantum shortcuts to adiabaticity for scale-invariant
  driving},}\ }\href {\doibase 10.1103/PhysRevX.4.021013} {\bibfield  {journal}
  {\bibinfo  {journal} {Phys. Rev. X}\ }\textbf {\bibinfo {volume} {4}},\
  \bibinfo {pages} {021013} (\bibinfo {year} {2014})}\BibitemShut {NoStop}%
\bibitem [{\citenamefont {Sels}\ and\ \citenamefont
  {Polkovnikov}(2017)}]{PNAS_Sels}%
  \BibitemOpen
  \bibfield  {author} {\bibinfo {author} {\bibfnamefont {Dries}\ \bibnamefont
  {Sels}}\ and\ \bibinfo {author} {\bibfnamefont {Anatoli}\ \bibnamefont
  {Polkovnikov}},\ }\bibfield  {title} {\enquote {\bibinfo {title} {Minimizing
  irreversible losses in quantum systems by local counterdiabatic driving},}\
  }\href {\doibase 10.1073/pnas.1619826114} {\bibfield  {journal} {\bibinfo
  {journal} {Proceedings of the National Academy of Sciences}\ }\textbf
  {\bibinfo {volume} {114}},\ \bibinfo {pages} {E3909--E3916} (\bibinfo {year}
  {2017})},\ \Eprint
  {http://arxiv.org/abs/https://www.pnas.org/content/114/20/E3909.full.pdf}
  {https://www.pnas.org/content/114/20/E3909.full.pdf} \BibitemShut {NoStop}%
\bibitem [{\citenamefont {Hartmann}\ \emph
  {et~al.}(2020{\natexlab{a}})\citenamefont {Hartmann}, \citenamefont
  {Mukherjee}, \citenamefont {Niedenzu},\ and\ \citenamefont
  {Lechner}}]{PRR_Wolfgang}%
  \BibitemOpen
  \bibfield  {author} {\bibinfo {author} {\bibfnamefont {Andreas}\ \bibnamefont
  {Hartmann}}, \bibinfo {author} {\bibfnamefont {Victor}\ \bibnamefont
  {Mukherjee}}, \bibinfo {author} {\bibfnamefont {Wolfgang}\ \bibnamefont
  {Niedenzu}}, \ and\ \bibinfo {author} {\bibfnamefont {Wolfgang}\ \bibnamefont
  {Lechner}},\ }\bibfield  {title} {\enquote {\bibinfo {title} {Many-body
  quantum heat engines with shortcuts to adiabaticity},}\ }\href {\doibase
  10.1103/PhysRevResearch.2.023145} {\bibfield  {journal} {\bibinfo  {journal}
  {Phys. Rev. Research}\ }\textbf {\bibinfo {volume} {2}},\ \bibinfo {pages}
  {023145} (\bibinfo {year} {2020}{\natexlab{a}})}\BibitemShut {NoStop}%
\bibitem [{\citenamefont {Zheng}\ \emph {et~al.}(2016)\citenamefont {Zheng},
  \citenamefont {Campbell}, \citenamefont {De~Chiara},\ and\ \citenamefont
  {Poletti}}]{PRA_Zhang}%
  \BibitemOpen
  \bibfield  {author} {\bibinfo {author} {\bibfnamefont {Yuanjian}\
  \bibnamefont {Zheng}}, \bibinfo {author} {\bibfnamefont {Steve}\ \bibnamefont
  {Campbell}}, \bibinfo {author} {\bibfnamefont {Gabriele}\ \bibnamefont
  {De~Chiara}}, \ and\ \bibinfo {author} {\bibfnamefont {Dario}\ \bibnamefont
  {Poletti}},\ }\bibfield  {title} {\enquote {\bibinfo {title} {Cost of
  counterdiabatic driving and work output},}\ }\href {\doibase
  10.1103/PhysRevA.94.042132} {\bibfield  {journal} {\bibinfo  {journal} {Phys.
  Rev. A}\ }\textbf {\bibinfo {volume} {94}},\ \bibinfo {pages} {042132}
  (\bibinfo {year} {2016})}\BibitemShut {NoStop}%
\bibitem [{\citenamefont {Calzetta}(2018)}]{PRA_Calzetta}%
  \BibitemOpen
  \bibfield  {author} {\bibinfo {author} {\bibfnamefont {Esteban}\ \bibnamefont
  {Calzetta}},\ }\bibfield  {title} {\enquote {\bibinfo {title} {Not-quite-free
  shortcuts to adiabaticity},}\ }\href {\doibase 10.1103/PhysRevA.98.032107}
  {\bibfield  {journal} {\bibinfo  {journal} {Phys. Rev. A}\ }\textbf {\bibinfo
  {volume} {98}},\ \bibinfo {pages} {032107} (\bibinfo {year}
  {2018})}\BibitemShut {NoStop}%
\bibitem [{\citenamefont {Torrontegui}\ \emph {et~al.}(2017)\citenamefont
  {Torrontegui}, \citenamefont {Lizuain}, \citenamefont {Gonz\'alez-Resines},
  \citenamefont {Tobalina}, \citenamefont {Ruschhaupt}, \citenamefont
  {Kosloff},\ and\ \citenamefont {Muga}}]{PRA_Muga}%
  \BibitemOpen
  \bibfield  {author} {\bibinfo {author} {\bibfnamefont {E.}~\bibnamefont
  {Torrontegui}}, \bibinfo {author} {\bibfnamefont {I.}~\bibnamefont
  {Lizuain}}, \bibinfo {author} {\bibfnamefont {S.}~\bibnamefont
  {Gonz\'alez-Resines}}, \bibinfo {author} {\bibfnamefont {A.}~\bibnamefont
  {Tobalina}}, \bibinfo {author} {\bibfnamefont {A.}~\bibnamefont
  {Ruschhaupt}}, \bibinfo {author} {\bibfnamefont {R.}~\bibnamefont {Kosloff}},
  \ and\ \bibinfo {author} {\bibfnamefont {J.~G.}\ \bibnamefont {Muga}},\
  }\bibfield  {title} {\enquote {\bibinfo {title} {Energy consumption for
  shortcuts to adiabaticity},}\ }\href {\doibase 10.1103/PhysRevA.96.022133}
  {\bibfield  {journal} {\bibinfo  {journal} {Phys. Rev. A}\ }\textbf {\bibinfo
  {volume} {96}},\ \bibinfo {pages} {022133} (\bibinfo {year}
  {2017})}\BibitemShut {NoStop}%
\bibitem [{\citenamefont {Funo}\ \emph {et~al.}(2017)\citenamefont {Funo},
  \citenamefont {Zhang}, \citenamefont {Chatou}, \citenamefont {Kim},
  \citenamefont {Ueda},\ and\ \citenamefont {del Campo}}]{PRL_Funo}%
  \BibitemOpen
  \bibfield  {author} {\bibinfo {author} {\bibfnamefont {Ken}\ \bibnamefont
  {Funo}}, \bibinfo {author} {\bibfnamefont {Jing-Ning}\ \bibnamefont {Zhang}},
  \bibinfo {author} {\bibfnamefont {Cyril}\ \bibnamefont {Chatou}}, \bibinfo
  {author} {\bibfnamefont {Kihwan}\ \bibnamefont {Kim}}, \bibinfo {author}
  {\bibfnamefont {Masahito}\ \bibnamefont {Ueda}}, \ and\ \bibinfo {author}
  {\bibfnamefont {Adolfo}\ \bibnamefont {del Campo}},\ }\bibfield  {title}
  {\enquote {\bibinfo {title} {Universal work fluctuations during shortcuts to
  adiabaticity by counterdiabatic driving},}\ }\href {\doibase
  10.1103/PhysRevLett.118.100602} {\bibfield  {journal} {\bibinfo  {journal}
  {Phys. Rev. Lett.}\ }\textbf {\bibinfo {volume} {118}},\ \bibinfo {pages}
  {100602} (\bibinfo {year} {2017})}\BibitemShut {NoStop}%
\bibitem [{\citenamefont {Jaramillo}\ \emph {et~al.}(2016)\citenamefont
  {Jaramillo}, \citenamefont {Beau},\ and\ \citenamefont {del
  Campo}}]{NJP_Jaramillo}%
  \BibitemOpen
  \bibfield  {author} {\bibinfo {author} {\bibfnamefont {J}~\bibnamefont
  {Jaramillo}}, \bibinfo {author} {\bibfnamefont {M}~\bibnamefont {Beau}}, \
  and\ \bibinfo {author} {\bibfnamefont {A}~\bibnamefont {del Campo}},\
  }\bibfield  {title} {\enquote {\bibinfo {title} {Quantum supremacy of
  many-particle thermal machines},}\ }\href
  {http://stacks.iop.org/1367-2630/18/i=7/a=075019} {\bibfield  {journal}
  {\bibinfo  {journal} {New Journal of Physics}\ }\textbf {\bibinfo {volume}
  {18}},\ \bibinfo {pages} {075019} (\bibinfo {year} {2016})}\BibitemShut
  {NoStop}%
\bibitem [{\citenamefont {Altintas}\ and\ \citenamefont {M\"ustecapl\ifmmode
  \imath \else \i \fi{}o\ifmmode~\breve{g}\else
  \u{g}\fi{}lu}(2015)}]{PRE_Ferdi}%
  \BibitemOpen
  \bibfield  {author} {\bibinfo {author} {\bibfnamefont {Ferdi}\ \bibnamefont
  {Altintas}}\ and\ \bibinfo {author} {\bibfnamefont {\"Ozg\"ur~E.}\
  \bibnamefont {M\"ustecapl\ifmmode \imath \else \i
  \fi{}o\ifmmode~\breve{g}\else \u{g}\fi{}lu}},\ }\bibfield  {title} {\enquote
  {\bibinfo {title} {General formalism of local thermodynamics with an example:
  Quantum otto engine with a spin-$1/2$ coupled to an arbitrary spin},}\ }\href
  {\doibase 10.1103/PhysRevE.92.022142} {\bibfield  {journal} {\bibinfo
  {journal} {Phys. Rev. E}\ }\textbf {\bibinfo {volume} {92}},\ \bibinfo
  {pages} {022142} (\bibinfo {year} {2015})}\BibitemShut {NoStop}%
\bibitem [{\citenamefont {Altintas}\ \emph {et~al.}(2014)\citenamefont
  {Altintas}, \citenamefont {Hardal},\ and\ \citenamefont {M\"ustecapl\ifmmode
  \imath \else \i \fi{}o\ifmmode~\breve{g}\else \u{g}\fi{}lu}}]{PRE_Ferdi2}%
  \BibitemOpen
  \bibfield  {author} {\bibinfo {author} {\bibfnamefont {Ferdi}\ \bibnamefont
  {Altintas}}, \bibinfo {author} {\bibfnamefont {Ali \"U.~C.}\ \bibnamefont
  {Hardal}}, \ and\ \bibinfo {author} {\bibfnamefont {\"Ozg\"ur~E.}\
  \bibnamefont {M\"ustecapl\ifmmode \imath \else \i
  \fi{}o\ifmmode~\breve{g}\else \u{g}\fi{}lu}},\ }\bibfield  {title} {\enquote
  {\bibinfo {title} {Quantum correlated heat engine with spin squeezing},}\
  }\href {\doibase 10.1103/PhysRevE.90.032102} {\bibfield  {journal} {\bibinfo
  {journal} {Phys. Rev. E}\ }\textbf {\bibinfo {volume} {90}},\ \bibinfo
  {pages} {032102} (\bibinfo {year} {2014})}\BibitemShut {NoStop}%
\bibitem [{\citenamefont {{\c{C}}akmak}\ \emph {et~al.}(2016)\citenamefont
  {{\c{C}}akmak}, \citenamefont {Altintas},\ and\ \citenamefont
  {E.~M{\"u}stecapl{\i}o{\u{g}}lu}}]{EPJP_Selcuk}%
  \BibitemOpen
  \bibfield  {author} {\bibinfo {author} {\bibfnamefont {Sel{\c{c}}uk}\
  \bibnamefont {{\c{C}}akmak}}, \bibinfo {author} {\bibfnamefont {Ferdi}\
  \bibnamefont {Altintas}}, \ and\ \bibinfo {author} {\bibfnamefont
  {{\"O}zg{\"u}r}\ \bibnamefont {E.~M{\"u}stecapl{\i}o{\u{g}}lu}},\ }\bibfield
  {title} {\enquote {\bibinfo {title} {Lipkin-meshkov-glick model in a quantum
  otto cycle},}\ }\href {\doibase 10.1140/epjp/i2016-16197-0} {\bibfield
  {journal} {\bibinfo  {journal} {The European Physical Journal Plus}\ }\textbf
  {\bibinfo {volume} {131}},\ \bibinfo {pages} {197} (\bibinfo {year}
  {2016})}\BibitemShut {NoStop}%
\bibitem [{\citenamefont {Hewgill}\ \emph {et~al.}(2018)\citenamefont
  {Hewgill}, \citenamefont {Ferraro},\ and\ \citenamefont
  {De~Chiara}}]{PRA_Gabriele}%
  \BibitemOpen
  \bibfield  {author} {\bibinfo {author} {\bibfnamefont {Adam}\ \bibnamefont
  {Hewgill}}, \bibinfo {author} {\bibfnamefont {Alessandro}\ \bibnamefont
  {Ferraro}}, \ and\ \bibinfo {author} {\bibfnamefont {Gabriele}\ \bibnamefont
  {De~Chiara}},\ }\bibfield  {title} {\enquote {\bibinfo {title} {Quantum
  correlations and thermodynamic performances of two-qubit engines with local
  and common baths},}\ }\href {\doibase 10.1103/PhysRevA.98.042102} {\bibfield
  {journal} {\bibinfo  {journal} {Phys. Rev. A}\ }\textbf {\bibinfo {volume}
  {98}},\ \bibinfo {pages} {042102} (\bibinfo {year} {2018})}\BibitemShut
  {NoStop}%
\bibitem [{\citenamefont {T{\"u}rkpen{\c{c}}e}\ and\ \citenamefont
  {Altintas}(2019)}]{QIP_Turkpence}%
  \BibitemOpen
  \bibfield  {author} {\bibinfo {author} {\bibfnamefont {Deniz}\ \bibnamefont
  {T{\"u}rkpen{\c{c}}e}}\ and\ \bibinfo {author} {\bibfnamefont {Ferdi}\
  \bibnamefont {Altintas}},\ }\bibfield  {title} {\enquote {\bibinfo {title}
  {Coupled quantum otto heat engine and refrigerator with inner friction},}\
  }\href {\doibase 10.1007/s11128-019-2366-7} {\bibfield  {journal} {\bibinfo
  {journal} {Quantum Information Processing}\ }\textbf {\bibinfo {volume}
  {18}},\ \bibinfo {pages} {255} (\bibinfo {year} {2019})}\BibitemShut
  {NoStop}%
\bibitem [{\citenamefont {B.~S}\ \emph {et~al.}(2020)\citenamefont {B.~S},
  \citenamefont {Mukherjee}, \citenamefont {Divakaran},\ and\ \citenamefont
  {del Campo}}]{PRR_Revathy}%
  \BibitemOpen
  \bibfield  {author} {\bibinfo {author} {\bibfnamefont {Revathy}\ \bibnamefont
  {B.~S}}, \bibinfo {author} {\bibfnamefont {Victor}\ \bibnamefont
  {Mukherjee}}, \bibinfo {author} {\bibfnamefont {Uma}\ \bibnamefont
  {Divakaran}}, \ and\ \bibinfo {author} {\bibfnamefont {Adolfo}\ \bibnamefont
  {del Campo}},\ }\bibfield  {title} {\enquote {\bibinfo {title} {Universal
  finite-time thermodynamics of many-body quantum machines from kibble-zurek
  scaling},}\ }\href {\doibase 10.1103/PhysRevResearch.2.043247} {\bibfield
  {journal} {\bibinfo  {journal} {Phys. Rev. Research}\ }\textbf {\bibinfo
  {volume} {2}},\ \bibinfo {pages} {043247} (\bibinfo {year}
  {2020})}\BibitemShut {NoStop}%
\bibitem [{\citenamefont {Hartmann}\ \emph
  {et~al.}(2020{\natexlab{b}})\citenamefont {Hartmann}, \citenamefont
  {Mukherjee}, \citenamefont {Mbeng}, \citenamefont {Niedenzu},\ and\
  \citenamefont {Lechner}}]{arXiv_Wolfgang}%
  \BibitemOpen
  \bibfield  {author} {\bibinfo {author} {\bibfnamefont {Andreas}\ \bibnamefont
  {Hartmann}}, \bibinfo {author} {\bibfnamefont {Victor}\ \bibnamefont
  {Mukherjee}}, \bibinfo {author} {\bibfnamefont {Glen~Bigan}\ \bibnamefont
  {Mbeng}}, \bibinfo {author} {\bibfnamefont {Wolfgang}\ \bibnamefont
  {Niedenzu}}, \ and\ \bibinfo {author} {\bibfnamefont {Wolfgang}\ \bibnamefont
  {Lechner}},\ }\bibfield  {title} {\enquote {\bibinfo {title} {Multi-spin
  counter-diabatic driving in many-body quantum {O}tto refrigerators},}\ }\href
  {\doibase 10.22331/q-2020-12-24-377} {\bibfield  {journal} {\bibinfo
  {journal} {{Quantum}}\ }\textbf {\bibinfo {volume} {4}},\ \bibinfo {pages}
  {377} (\bibinfo {year} {2020}{\natexlab{b}})}\BibitemShut {NoStop}%
\bibitem [{\citenamefont {Takahashi}(2013)}]{PRE_Takahashi}%
  \BibitemOpen
  \bibfield  {author} {\bibinfo {author} {\bibfnamefont {Kazutaka}\
  \bibnamefont {Takahashi}},\ }\bibfield  {title} {\enquote {\bibinfo {title}
  {Transitionless quantum driving for spin systems},}\ }\href {\doibase
  10.1103/PhysRevE.87.062117} {\bibfield  {journal} {\bibinfo  {journal} {Phys.
  Rev. E}\ }\textbf {\bibinfo {volume} {87}},\ \bibinfo {pages} {062117}
  (\bibinfo {year} {2013})}\BibitemShut {NoStop}%
\bibitem [{\citenamefont {del Campo}\ \emph {et~al.}(2012)\citenamefont {del
  Campo}, \citenamefont {Rams},\ and\ \citenamefont
  {Zurek}}]{PRL_delCampoIsing}%
  \BibitemOpen
  \bibfield  {author} {\bibinfo {author} {\bibfnamefont {Adolfo}\ \bibnamefont
  {del Campo}}, \bibinfo {author} {\bibfnamefont {Marek~M.}\ \bibnamefont
  {Rams}}, \ and\ \bibinfo {author} {\bibfnamefont {Wojciech~H.}\ \bibnamefont
  {Zurek}},\ }\bibfield  {title} {\enquote {\bibinfo {title} {Assisted
  finite-rate adiabatic passage across a quantum critical point: Exact solution
  for the quantum ising model},}\ }\href {\doibase
  10.1103/PhysRevLett.109.115703} {\bibfield  {journal} {\bibinfo  {journal}
  {Phys. Rev. Lett.}\ }\textbf {\bibinfo {volume} {109}},\ \bibinfo {pages}
  {115703} (\bibinfo {year} {2012})}\BibitemShut {NoStop}%
\bibitem [{\citenamefont {Ib\'a\~nez}\ \emph {et~al.}(2012)\citenamefont
  {Ib\'a\~nez}, \citenamefont {Chen}, \citenamefont {Torrontegui},
  \citenamefont {Muga},\ and\ \citenamefont {Ruschhaupt}}]{PRL_Ibanez}%
  \BibitemOpen
  \bibfield  {author} {\bibinfo {author} {\bibfnamefont {S.}~\bibnamefont
  {Ib\'a\~nez}}, \bibinfo {author} {\bibfnamefont {Xi}~\bibnamefont {Chen}},
  \bibinfo {author} {\bibfnamefont {E.}~\bibnamefont {Torrontegui}}, \bibinfo
  {author} {\bibfnamefont {J.~G.}\ \bibnamefont {Muga}}, \ and\ \bibinfo
  {author} {\bibfnamefont {A.}~\bibnamefont {Ruschhaupt}},\ }\bibfield  {title}
  {\enquote {\bibinfo {title} {Multiple schr\"odinger pictures and dynamics in
  shortcuts to adiabaticity},}\ }\href {\doibase
  10.1103/PhysRevLett.109.100403} {\bibfield  {journal} {\bibinfo  {journal}
  {Phys. Rev. Lett.}\ }\textbf {\bibinfo {volume} {109}},\ \bibinfo {pages}
  {100403} (\bibinfo {year} {2012})}\BibitemShut {NoStop}%
\bibitem [{\citenamefont {Stefanatos}\ and\ \citenamefont
  {Paspalakis}(2019)}]{PRA_Stefanatos}%
  \BibitemOpen
  \bibfield  {author} {\bibinfo {author} {\bibfnamefont {Dionisis}\
  \bibnamefont {Stefanatos}}\ and\ \bibinfo {author} {\bibfnamefont {Emmanuel}\
  \bibnamefont {Paspalakis}},\ }\bibfield  {title} {\enquote {\bibinfo {title}
  {Efficient generation of the triplet bell state between coupled spins using
  transitionless quantum driving and optimal control},}\ }\href {\doibase
  10.1103/PhysRevA.99.022327} {\bibfield  {journal} {\bibinfo  {journal} {Phys.
  Rev. A}\ }\textbf {\bibinfo {volume} {99}},\ \bibinfo {pages} {022327}
  (\bibinfo {year} {2019})}\BibitemShut {NoStop}%
\bibitem [{\citenamefont {Deffner}\ and\ \citenamefont
  {Lutz}(2010)}]{PRL_Deffner}%
  \BibitemOpen
  \bibfield  {author} {\bibinfo {author} {\bibfnamefont {Sebastian}\
  \bibnamefont {Deffner}}\ and\ \bibinfo {author} {\bibfnamefont {Eric}\
  \bibnamefont {Lutz}},\ }\bibfield  {title} {\enquote {\bibinfo {title}
  {Generalized clausius inequality for nonequilibrium quantum processes},}\
  }\href {\doibase 10.1103/PhysRevLett.105.170402} {\bibfield  {journal}
  {\bibinfo  {journal} {Phys. Rev. Lett.}\ }\textbf {\bibinfo {volume} {105}},\
  \bibinfo {pages} {170402} (\bibinfo {year} {2010})}\BibitemShut {NoStop}%
\bibitem [{\citenamefont {Plastina}\ \emph {et~al.}(2014)\citenamefont
  {Plastina}, \citenamefont {Alecce}, \citenamefont {Apollaro}, \citenamefont
  {Falcone}, \citenamefont {Francica}, \citenamefont {Galve}, \citenamefont
  {Lo~Gullo},\ and\ \citenamefont {Zambrini}}]{PRL_Plastina}%
  \BibitemOpen
  \bibfield  {author} {\bibinfo {author} {\bibfnamefont {F.}~\bibnamefont
  {Plastina}}, \bibinfo {author} {\bibfnamefont {A.}~\bibnamefont {Alecce}},
  \bibinfo {author} {\bibfnamefont {T.~J.~G.}\ \bibnamefont {Apollaro}},
  \bibinfo {author} {\bibfnamefont {G.}~\bibnamefont {Falcone}}, \bibinfo
  {author} {\bibfnamefont {G.}~\bibnamefont {Francica}}, \bibinfo {author}
  {\bibfnamefont {F.}~\bibnamefont {Galve}}, \bibinfo {author} {\bibfnamefont
  {N.}~\bibnamefont {Lo~Gullo}}, \ and\ \bibinfo {author} {\bibfnamefont
  {R.}~\bibnamefont {Zambrini}},\ }\bibfield  {title} {\enquote {\bibinfo
  {title} {Irreversible work and inner friction in quantum thermodynamic
  processes},}\ }\href {\doibase 10.1103/PhysRevLett.113.260601} {\bibfield
  {journal} {\bibinfo  {journal} {Phys. Rev. Lett.}\ }\textbf {\bibinfo
  {volume} {113}},\ \bibinfo {pages} {260601} (\bibinfo {year}
  {2014})}\BibitemShut {NoStop}%
\bibitem [{\citenamefont {Feldmann}\ and\ \citenamefont
  {Kosloff}(2003)}]{PRE_Feldmann}%
  \BibitemOpen
  \bibfield  {author} {\bibinfo {author} {\bibfnamefont {Tova}\ \bibnamefont
  {Feldmann}}\ and\ \bibinfo {author} {\bibfnamefont {Ronnie}\ \bibnamefont
  {Kosloff}},\ }\bibfield  {title} {\enquote {\bibinfo {title} {Quantum
  four-stroke heat engine: Thermodynamic observables in a model with intrinsic
  friction},}\ }\href {\doibase 10.1103/PhysRevE.68.016101} {\bibfield
  {journal} {\bibinfo  {journal} {Phys. Rev. E}\ }\textbf {\bibinfo {volume}
  {68}},\ \bibinfo {pages} {016101} (\bibinfo {year} {2003})}\BibitemShut
  {NoStop}%
\bibitem [{\citenamefont {Alecce}\ \emph {et~al.}(2015)\citenamefont {Alecce},
  \citenamefont {Galve}, \citenamefont {Gullo}, \citenamefont {Dell'Anna},
  \citenamefont {Plastina},\ and\ \citenamefont {Zambrini}}]{NJP_Alecce}%
  \BibitemOpen
  \bibfield  {author} {\bibinfo {author} {\bibfnamefont {A}~\bibnamefont
  {Alecce}}, \bibinfo {author} {\bibfnamefont {F}~\bibnamefont {Galve}},
  \bibinfo {author} {\bibfnamefont {N~Lo}\ \bibnamefont {Gullo}}, \bibinfo
  {author} {\bibfnamefont {L}~\bibnamefont {Dell'Anna}}, \bibinfo {author}
  {\bibfnamefont {F}~\bibnamefont {Plastina}}, \ and\ \bibinfo {author}
  {\bibfnamefont {R}~\bibnamefont {Zambrini}},\ }\bibfield  {title} {\enquote
  {\bibinfo {title} {Quantum otto cycle with inner friction: finite-time and
  disorder effects},}\ }\href {\doibase 10.1088/1367-2630/17/7/075007}
  {\bibfield  {journal} {\bibinfo  {journal} {New Journal of Physics}\ }\textbf
  {\bibinfo {volume} {17}},\ \bibinfo {pages} {075007} (\bibinfo {year}
  {2015})}\BibitemShut {NoStop}%
\bibitem [{\citenamefont {{\c{C}}akmak}\ \emph {et~al.}(2017)\citenamefont
  {{\c{C}}akmak}, \citenamefont {Altintas}, \citenamefont {Gen{\c{c}}ten},\
  and\ \citenamefont {M{\"u}stecapl{\i}o{\u{g}}lu}}]{EPJD_Selcuk}%
  \BibitemOpen
  \bibfield  {author} {\bibinfo {author} {\bibfnamefont {Sel{\c{c}}uk}\
  \bibnamefont {{\c{C}}akmak}}, \bibinfo {author} {\bibfnamefont {Ferdi}\
  \bibnamefont {Altintas}}, \bibinfo {author} {\bibfnamefont {Azmi}\
  \bibnamefont {Gen{\c{c}}ten}}, \ and\ \bibinfo {author} {\bibfnamefont
  {{\"O}zg{\"u}r~E.}\ \bibnamefont {M{\"u}stecapl{\i}o{\u{g}}lu}},\ }\bibfield
  {title} {\enquote {\bibinfo {title} {Irreversible work and internal friction
  in a quantum otto cycle of a single arbitrary spin},}\ }\href {\doibase
  10.1140/epjd/e2017-70443-1} {\bibfield  {journal} {\bibinfo  {journal} {The
  European Physical Journal D}\ }\textbf {\bibinfo {volume} {71}},\ \bibinfo
  {pages} {75} (\bibinfo {year} {2017})}\BibitemShut {NoStop}%
\bibitem [{\citenamefont {Feldmann}\ and\ \citenamefont
  {Kosloff}(2006)}]{PRE_Feldmann2}%
  \BibitemOpen
  \bibfield  {author} {\bibinfo {author} {\bibfnamefont {Tova}\ \bibnamefont
  {Feldmann}}\ and\ \bibinfo {author} {\bibfnamefont {Ronnie}\ \bibnamefont
  {Kosloff}},\ }\bibfield  {title} {\enquote {\bibinfo {title} {Quantum
  lubrication: Suppression of friction in a first-principles four-stroke heat
  engine},}\ }\href {\doibase 10.1103/PhysRevE.73.025107} {\bibfield  {journal}
  {\bibinfo  {journal} {Phys. Rev. E}\ }\textbf {\bibinfo {volume} {73}},\
  \bibinfo {pages} {025107} (\bibinfo {year} {2006})}\BibitemShut {NoStop}%
\bibitem [{\citenamefont {Rezek}\ and\ \citenamefont
  {Kosloff}(2006)}]{NJP_Rezek}%
  \BibitemOpen
  \bibfield  {author} {\bibinfo {author} {\bibfnamefont {Yair}\ \bibnamefont
  {Rezek}}\ and\ \bibinfo {author} {\bibfnamefont {Ronnie}\ \bibnamefont
  {Kosloff}},\ }\bibfield  {title} {\enquote {\bibinfo {title} {Irreversible
  performance of a quantum harmonic heat engine},}\ }\href
  {http://stacks.iop.org/1367-2630/8/i=5/a=083} {\bibfield  {journal} {\bibinfo
   {journal} {New Journal of Physics}\ }\textbf {\bibinfo {volume} {8}},\
  \bibinfo {pages} {83} (\bibinfo {year} {2006})}\BibitemShut {NoStop}%
\bibitem [{\citenamefont {{\c{C}}akmak}\ \emph {et~al.}(2020)\citenamefont
  {{\c{C}}akmak}, \citenamefont {{\c{C}}and{\i}r},\ and\ \citenamefont
  {Altintas}}]{Ferdi_lags}%
  \BibitemOpen
  \bibfield  {author} {\bibinfo {author} {\bibfnamefont {Sel{\c{c}}uk}\
  \bibnamefont {{\c{C}}akmak}}, \bibinfo {author} {\bibfnamefont {Mustafa}\
  \bibnamefont {{\c{C}}and{\i}r}}, \ and\ \bibinfo {author} {\bibfnamefont
  {Ferdi}\ \bibnamefont {Altintas}},\ }\bibfield  {title} {\enquote {\bibinfo
  {title} {Construction of a quantum carnot heat engine cycle},}\ }\href
  {\doibase 10.1007/s11128-020-02831-1} {\bibfield  {journal} {\bibinfo
  {journal} {Quantum Information Processing}\ }\textbf {\bibinfo {volume}
  {19}},\ \bibinfo {pages} {314} (\bibinfo {year} {2020})}\BibitemShut
  {NoStop}%
\bibitem [{\citenamefont {Singh}\ and\ \citenamefont
  {Abah}(2020)}]{arXiv_Singh}%
  \BibitemOpen
  \bibfield  {author} {\bibinfo {author} {\bibfnamefont {Satnam}\ \bibnamefont
  {Singh}}\ and\ \bibinfo {author} {\bibfnamefont {Obinna}\ \bibnamefont
  {Abah}},\ }\bibfield  {title} {\enquote {\bibinfo {title} {Energy
  optimization of two-level quantum otto machines},}\ }\href
  {https://arxiv.org/abs/2008.05002} {\bibfield  {journal} {\bibinfo  {journal}
  {arXiv:2008.05002}\ } (\bibinfo {year} {2020})},\ \Eprint
  {http://arxiv.org/abs/2008.05002} {arXiv:2008.05002 [cond-mat.stat-mech]}
  \BibitemShut {NoStop}%
\bibitem [{\citenamefont {Sgroi}\ \emph {et~al.}(2021)\citenamefont {Sgroi},
  \citenamefont {Palma},\ and\ \citenamefont {Paternostro}}]{PRL_Sgroi}%
  \BibitemOpen
  \bibfield  {author} {\bibinfo {author} {\bibfnamefont {Pierpaolo}\
  \bibnamefont {Sgroi}}, \bibinfo {author} {\bibfnamefont {G.~Massimo}\
  \bibnamefont {Palma}}, \ and\ \bibinfo {author} {\bibfnamefont {Mauro}\
  \bibnamefont {Paternostro}},\ }\bibfield  {title} {\enquote {\bibinfo {title}
  {Reinforcement learning approach to nonequilibrium quantum thermodynamics},}\
  }\href {\doibase 10.1103/PhysRevLett.126.020601} {\bibfield  {journal}
  {\bibinfo  {journal} {Phys. Rev. Lett.}\ }\textbf {\bibinfo {volume} {126}},\
  \bibinfo {pages} {020601} (\bibinfo {year} {2021})}\BibitemShut {NoStop}%
\end{thebibliography}%

% new refs: arXiv_Mukherjee, Entropy_Beau, NJP_Jaramillo, PRR_Revathy, PRL_Funo, PRL_delCampoIsing, PRA_Deng, Demirplak2005, Demirplak2008

\end{document}